# The Effect of Spike Geometry on the Linear and Nonlinear Plasmonic Properties of Gold Nanourchins

Domantas Peckus[1,2*], Fatima Albatoul Kasabji[3,4], Maziar Moussavi[1], Loic Vidal[3,4], Hana Boubaker[3,4], Asta Tamulevičienė[1,2], Arnaud Spangenberg[3,4], Tomas Tamulevičius[1,2], Joel Henzie[5*], Karine Mougin[3,4*], Sigitas Tamulevičius[1,2*]

[1]Institute of Materials Science of Kaunas University of Technology, K. Baršausko st. 59, LT-51423, Kaunas, Lithuania

[2]Department of Physics, Kaunas University of Technology, Studentų st. 50, LT-51368, Kaunas, Lithuania

[3]Institut de Science des Matériaux de Mulhouse IS2M UMR 7361, 15 rue Jean Starcky, F 68100 Mulhouse, France

[4]Université de Strasbourg, 4 Rue Blaise Pascal, CS 90032, F-67081 Strasbourg, France

[5]International Center for Materials Nanoarchitectonics (WPI-MANA), National Institute for Materials Science (NIMS), Tsukuba, Ibaraki 305-0044, Japan

*Corresponding authors

D. Peckus: domantas.peckus@ktu.lt, orcid.org/0000-0002-4224-2521;

J. Henzie: henzie.joeladam@nims.go.jp, orcid.org/0000-0002-9190-2645;

K. Mougin: karine.mougin@uha.fr, orcid.org/0000-0002-0481-1594;

S. Tamulevičius: sigitas.tamulevicius@ktu.lt, orcid.org/0000-0002-9965-2724.

**Abstract**

Wet-chemistry synthesized Gold nanourchins (Au NUs), characterized by spiky morphologies with spherical cores, exhibit complex and geometry-dependent plasmonic properties distinct from those of symmetrical nanostructures. While plasmon hybridization and mode coupling in branched nanostructures have been broadly studied, the specific optical behavior of Au NUs—particularly regarding spike length distribution and ultrafast dynamics—remains underexplored. This study investigates the steady-state and transient absorption spectra of Au NUs with 50–80 nm cores and 5–20 nm spikes, revealing multiple resonance bands. Transient absorption spectroscopy at various excitation wavelengths confirmed the presence of distinct resonances. Electromagnetic simulations based on TEM tomography-inspired models identified two key extinction bands: a green-wavelength dark mode resonance and a red/nIR spike-induced resonance attributed to the lightning rod effect. Simulations further showed that short, uniformly distributed spikes (aspect ratio ≤ 1) weakly excite dark resonances, while longer spikes (aspect ratio > 1) induce hybridized longitudinal resonances and significant redshifts. Broad spike length distributions result in multiple coexisting resonances, aligning with experimental extinction spectra. A preliminary surface-enhanced Raman scattering study using 2-naphthalene thiol confirmed stronger enhancement for long-spiked Au NUs under 785 nm excitation, validating the field enhancement potential of the identified resonances.

**Keywords:** Au nanourchins, spikes of Au nanourchins, localized surface plasmon resonance, transient absorption spectroscopy, EM modeling

## 1. Introduction

Plasmonic metallic nanoparticles and nanostructures are interesting due to their optoelectronic properties. Leveraging their size and shape-dependent localized surface plasmon resonance (LSPR) properties, nanostructures can be used for photocatalysis [1], nonlinear optics [2], surface-enhanced Raman spectroscopy (SERS) [3], enhanced fluorescence [4], sensors [5], photovoltaics [6], biosensors [7], and medicine [8,9]. Selecting the most suitable plasmonic material for a given nanostructure application necessitates consideration of factors such as the metal type [10], nanostructure size [11], shape [11], and crystallinity [12]. So far, the most popular metals for plasmonic applications are Au and Ag. Various forms of Au nanostructures can be synthesized, resulting in symmetrical forms like spheres, cubes, tetrahedra, and nanorods [11] or nonsymmetrical - like nanourchins [13,14], nanostars [4,15,16], and nanoflowers [17]. These nonsymetrical nanostructures can look quite similar; therefore, sometimes these names are used as synonyms [8], because both have core and spikes, but there are some characteristic differences too. Nanourchins have a spiky, spherical shape, similar to sea urchins, with many protrusive, sharp tips. Their surface is usually covered with small, uneven length and thickness spikes [18,19]. On the other hand, nanostars have a star-shaped morphology with a central core and several sharp, well-defined, elongated tips extending outwards [20–22]. While instead of spikes, Au nanoflowers have petals attached to the core [23,24].

The optical properties of symmetric Au nanoparticles are well described by Mie scattering theory, whereas the properties of asymmetric Au nanoparticles are less thoroughly explored, despite their potential for higher efficiency in specific applications such as medicine, biosensing, in vivo imaging, drug delivery, photothermal therapy, and photodynamic therapy [8]. Au nanostars, a counterpart of Au nanourchins, have already proven to be very efficient for SERS[20]. It has been demonstrated that Au nanourchins offer advantages over Au nanospheres or Au nanorods in SERS measurements [7,13,18]. Currently, more scientific research on optical properties is being conducted on Au nanostars compared to Au nanourchins [25,26], but both of them seem to be promising for future plasmonic applications.

Various wet-chemistry methods exist for the synthesis of Au nanourchins, but they can generally be classified into seeded or seedless growth approaches. Among these, the seed-mediated method, where silver nanoparticles serve as silver seeds, is the most employed for preparing Au nanourchins. By separating the nucleation and growth processes, this method provides better control over the final size and morphology of Au NUs. Factors such as solution pH, temperature, seed quantity, size, composition, and other variables can be adjusted to achieve the desired characteristics [27]. Silver seeds used to prepare Au nanourchins were successfully synthesized by F.G. Xu *et al.* using a seed-mediated method, which is quite facile and does not need any template or surfactant agent [28]. N.G. Bastús *et al.* showed the possibility of synthesizing highly monodisperse and long-term stable and functionalizable Ag NPs [29,30].

NPs results in various shape and high-resolution transmission electron microscopy (HR-TEM) or scanning electron microscopy (SEM) [31] is usually the first choice for inspection. On the other hand, the preparation of a sample for measurements takes a lot of time, and a limited number of nanoparticles are usually investigated in comparison to optical methods [32].

One of the most used indirect ways to determine the size, shape, and homogeneity of plasmonic metal nanoparticles is steady-state UV-Vis-nIR absorption measurements [14,32,33], where different geometrical form factors influence the light scattering and absorption properties of nanoparticles, resulting in different LSPRs. Machine learning-based methods for interpreting extinction spectra are emerging [34]. The rigorous numerical methods help to extend the understanding of the LSPRs in the optical extinction spectra of more complex nanostructure geometries than spheres and ellipsoids [34]. However, steady-state optical spectra can still hinder some hard-to-deconvolute information due to the overlapping resonances. In contrast, the transient absorption ultrafast optical measurements can reveal additional important information, revealing resonances hindered by the dominating ones, especially about more complicated nanostructures [11]. It was shown that transient absorption spectroscopy (TAS) provided more details about the shape of nanoparticles in comparison to steady-state absorption spectra, which is limited to the analysis of Mie resonance decomposition [11]. The auxiliary information on the LSPR decay dynamics, TAS spectral composition of ground state bleaching can be very advantageous for the analysis of asymmetric Au nanostructures that were not systematically studied with TAS yet. TAS can provide information about the different LSPR species that are too weak to be observed with steady-state spectroscopy, as it was shown in [11]. Recent ultrafast plasmonic property studies with TAS span over variously shaped Au nanoparticles [11], including spheres [11], nanorods [11,35], and nanoporous materials [36,37].

Despite growing interest in nanourchins resulting in 82 hits in Web of Knowledge, of which 46 are about Au nanostructures, their optical properties are undeservedly still not explained in detail. The complex hybridized resonances in Au nanourchins are known to be composed of and influenced by the spherical core and random length and density nanospikes/nanorods, which were analyzed separately [11]. Likewise, a comprehensive TAS analysis of Au nanourchins has not been conducted yet. A deeper understanding of the ultrafast properties of LSPR in Au nanourchins could significantly enhance their utility across diverse applications in photonics. Therefore, we devoted this work to the first deep analysis of steady state and dynamic plasmonic optical properties of Au nanourchins. The proposed approach could be extended for Au nanostars and Au nanoflowers.

In this work, we have synthesized a set of Au nanourchins with different core sizes (50-80 nm) and spike lengths (6-15 nm) along with spherical Au nanoparticles of the same size. The conducted TEM imaging and tomography studies inspired the use of electromagnetic (EM) simulations and helped to explain the nature of the emerging broad extinction in the optical spectrum. Two different plasmonic bands in Au nanourchins with long spikes were observed not met for the short ones and explained the significance of the higher aspect ratio spikes hindered in the steady state extinction spectroscopy. The SERS measurements confirmed the importance of the electric field concentration in the longer spike nanourchins met in the numerical models.

## 2 Experimental

### 2.1 Chemical Synthesis

*2.1.1 Materials*

**Chemicals.** Tannic acid (TA) ($C_{76}H_{52}O_{46}$), silver nitrate ($AgNO_3$), trisodium citrate ($Na_3C_6H_5O_7$), chloroauric acid ($HAuCl_4$), polyvinylpyrrolidone (PVP 40), and L-3,4-dihydroxyphenylalanine (L-DOPA), trisodium citrate dihydrate ($C_6H_5Na_3O_7 \cdot 2H_2O$) were purchased from Sigma-Aldrich. Ultrapure water ($\rho$ = 18.2 MΩ·cm) was used in all experiments. Glassware was rinsed before use with aqua regia and then with pure water.

*2.1.2 Synthesis of Gold Nanospheres*

Synthesis of Au nanospheres (Au NSs) for imitating the core of the nanourchins was performed according to the route described in [51]. In short, the synthesis consists of two steps, starting from seed synthesis and further growth of nanoparticles. Seeds were synthesized by heating 150 ml of 2.2 mM sodium citrate solution for 15 min under vigorous stirring. After the solution started to boil, 1 ml of $HAuCl_4$ (25 mM) was injected. The colour of the solution changed to soft pink in 10 min. Immediately after the seed synthesis, the reaction was cooled to 90°C, and 1 ml of $HAuCl_4$ (25 mM) was injected. The solution was left for 30 minutes stirring and heating at 90°C. These injection and stirring procedures were repeated two additional times. After that, 55 ml of the sample was extracted and added to the same vessel, 53 ml of water and 2 ml of 60 mM sodium citrate were added. This solution was then used as a seed solution for further growth. Repeating this sequence 4–6 times resulted in final nanoparticle diameters of 50–80 nm, respectively.

*2.1.3 Synthesis of Silver Seeds*

The silver nanoseeds were prepared according to the typical method reported by N.G. Bastús *et al.* They presented a method to synthesize a highly monodisperse sodium citrate-coated spherical silver nanoparticles (Ag NPs) by following a kinetically controlled seeded-growth approach via the reduction of silver nitrate by the combination of two chemical reducing agents: sodium citrate and tannic acid [29]. By adjusting the tannic acid concentration from 0.25 mM to 1 mM, the size of Ag NPs increased from 21.4 ± 5.0 nm to 37 ± 6.2 nm, respectively. The prepared silver seeds were used in the next section for the synthesis of gold nanourchins.

*2.1.4 Synthesis of Au Nanourchins*

The synthesis of gold nanourchins (Au NUs), was performed according to the formula proposed by A. Silvestri *et al.* [52]. Two solutions were prepared, solution "A", which is a mix of as-prepared silver seeds solution (see section 2.1.3), L-DOPA (12 mM). and ultrapure water with a 1:1:2.9 ratio, while solution "B" is $HAuCl_4$ solution (12 mM). The synthesis was carried out in a 'T'-shaped Mixer using two syringe pumps (**Figure S1**).

The microfluidic system was used to allow better mixing of the product and ensure homogeneity in the distribution of the NU's size. The core size was controlled by varying the concentration of tannic acid during the synthesis of the silver seeds, while the spike length was influenced by the timing of PVP addition during the synthesis of the nanourchins. More details are provided in the supporting information "Chemical synthesis" section (**Figure S2**).

A monolayer of gold nanoparticles was deposited on silicon according to the method described in [53]. 8 ml of colloidal solution (Au NSs or Au NUs) was diluted with 2 ml of deionized water in a glass beaker. Then, 8 ml of hexane was added dropwise on the surface of the solution. It forms an incompatible oil – water interface. At the end, 10 ml of ethanol is added slowly to the aqueous solution. The layer of NPs starts to assemble at the interface between the colloid-ethanol and hexane interface. After hexane evaporates, the monolayer is transferred on silicon.

2.2 Characterization

*2.2.1 Transmission electron microscopy (TEM)*

A drop of nanoparticles was dried on a carbon film grid and the micrographs were acquired with a high-resolution transmission electron microscope (HR-TEM) ARM200 (JEOL) equipped with a cold field emission gun (FEG) and used an acceleration voltage was 200 kV. Imaging using a rotating sample holder helped to determine the characteristic linear dimensions and size distributions of Au NUs and Au NSs. To achieve a three-dimensional visualization of the spikey particles, the nanourchins were imaged using TEM by tilting them from -50° to +60° in 1° increments, resulting in a total of 111 micrographs. These images were processed using a custom Python script, which employs the Local Feature Transformer (LoFTR) feature-matching algorithm [54] to align consecutive frames and correct for translational and rotational displacements. Dimensions of the NPs were measured with ImageJ software.

*2.2.2 UV-Visible Spectroscopy*

UV-Visible-near Infrared extinction spectra of the nanostructures under study were recorded using a fiber-optic spectrometer, AvaSpec-2048 (Avantes), with a resolution of 1.4 nm within the 200-1100 nm spectral range and utilizing a combined Deuterium Halogen light source, AvaLight-DHc (Avantes).

*2.2.3 Transient absorption spectroscopy*

Ultrafast relaxation processes in Au NPs (nanourchins, nanospheres) were investigated using a transient absorption spectrometer, HARPIA (Light Conversion). The system was excited with an ultrafast 290 fs pulse length and 1030 nm wavelength Yb:KGW laser Pharos (Light Conversion), with a regenerative amplifier at a 66.7 kHz repetition rate. The pump beam wavelength, based on the LSPR absorption maxima of the investigated NPs, was tuned to 350, 500, 600, 700, 750, and 800 nm using a collinear optical parametric generator, Orpheus, and a harmonic generator, Lyra (Light Conversion). The samples were subsequently probed with a white light supercontinuum generated using a 2 mm thickness sapphire plate excited with a fundamental laser wavelength (1030 nm). The spectral range of the supercontinuum probe, as well as the detection range of the TAS dynamics, spanned wavelengths from 480 to 783 nm. The excitation beam was focused on the 1 mm optical path length quartz cuvette (Hellma Analytics) to an approximately 700 μm diameter spot, overlaid with the supercontinuum probe, which was approximately 500 μm in diameter. The volume of the Au nanoparticle sample in water analyzed by TAS was approximately $2 \times 10^{-7}$ mm³.

*2.2.4 SERS analysis*

The Raman spectral data of representative nanosphere and nanourchin samples were acquired using a micro-Raman spectrometer inVia (Renishaw) equipped with a thermoelectrically cooled 2048-pixel CCD detector. Samples were excited with two lasers through a 50x/0.75 NA objective (Leica): 532 nm (excitation power was 0.36 mW, laser spot size ~1 mm, 2400 lines/mm diffraction grating) and 785 nm (excitation power was 0.136 mW, laser spot size ~2 mm, 1200 lines/mm diffraction grating). The signal was collected for 10 seconds. 2-Naphthalenethiol (2NT) of $10^{-4}$M concentration was used for testing the signal enhancement. Monolayer samples were immersed in the analyte solution and kept for 1h, afterwards, the samples were taken out and washed with ethanol to remove the excessive amount of analyte from the surface. The analytical enhancement factor was calculated by comparing the signal with the reference of the signal acquired using $10^{-2}$ M analyte concentration.

*2.2.5 Electromagnetic simulations*

Full-field EM simulations were performed on Au nanoparticles to estimate their optical properties at a wavelength range of 400 to 1200 nanometers when suspended in a medium equivalent to water ($n$ = 1.333). Nanourchin (NU) particles described in **Table S1** and **Section 3.1** were used as inspiration to create morphological models that possess the most important structural features using a code developed in Matlab. The models were then imported into the EM simulation software Lumerical. NUs with no core were created by overlapping their core using a sphere with $n$ = 1.333. The Au optical constants described in McPeak *et al.* [55] were used to model the optical response of all nanoparticles because it has good coverage from visible to near-infrared wavelengths. Each Au NU model was excited with a broadband total field scattered field source, then the scattered and absorbed light was recorded with frequency-domain power monitors enclosing the particles.

**3. Results and Discussion**

3.1 TEM image analysis

Aiming to understand the optical properties of complex Au nanostructures, representative spherical gold nanoparticle samples were used, resembling the core or the overall effective size of the gold nanourchins. The wet chemistry method synthesized Au NPs resulted in 52±5 nm, 63±6 nm, and 78±8 nm in diameter, as was obtained from the TEM micrographs (**Figure 1**). More complex Au nanostructures—nanourchins with varying core sizes and spike lengths—were synthesized using the seed-mediated wet-chemistry method with Ag seeds (**Figure 1, S3**). The synthesized nanostructure TEM micrograph analysis results are summarized in **Table 1** where samples are grouped based on their characteristic linear dimensions.

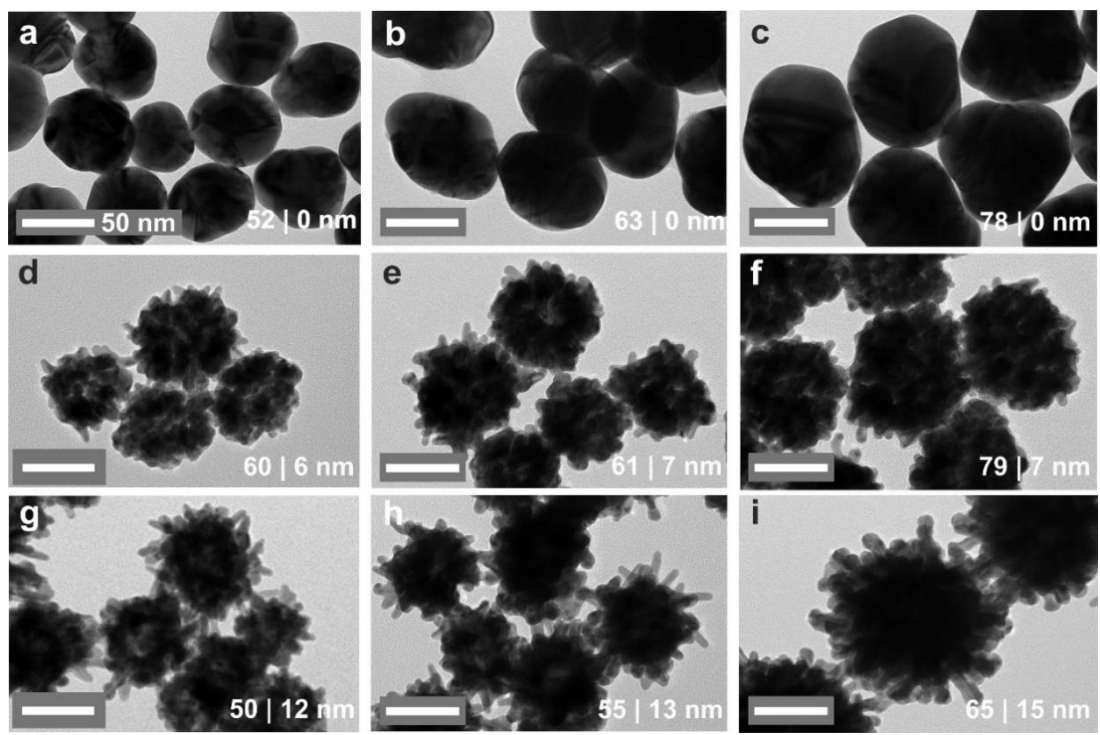

**Figure 1**. TEM images of Au nanospheres: Au NPs 52|0 nm (**a**); Au NPs 63|0 nm (**b**); Au NPs 78|0 nm (**c**) and Au nanourchins: Au NUs 60|6 nm (**d**); Au NUs 61|7 nm (**e**); Au NUs 79|7 nm (**f**); Au NUs 50|12 nm (**g**); Au NUs 55|13 nm (**h**) and Au NUs 65|15 nm (**i**). Three different groups of samples are in this figure first group **a**, **d**, **g**; the second **b**, **e**, **h** and the third **c**, **f**, **i**.

**Table 1**. Characteristic linear dimensions of Au nanosphere (NS) and Au nanourchins (NU). Their typical synthesis conditions, used Ag seed size, Au NS and Au NU core diameters (D), spike lengths (L), and their aspect ratios (L/D) as obtained from **Figures 1, 3S**.

| Sample | | Comment on synthesis specifics | Core\|D | Spike\|L | Spike\|D | Spike Aspect Ratio |
|---|---|---|---|---|---|---|
| Name | NP type | | | | | |
| 52\|00 | Au NS | 4 cycles: 53 ml of water and 2 ml of 60 mM sodium citrate, 25 mM HAuCl$_4$ | 51.5±4.7 | - | - | - |
| 63\|00 | | 5 cycles: 53 ml of water and 2 ml of 60 mM sodium citrate, 25 mM HAuCl$_4$ | 62.5±6.2 | - | - | - |
| 78\|00 | | 6 cycles: 53 ml of water and 2 ml of 60 mM sodium citrate, 25 mM HAuCl$_4$ | 77.5±7.9 | - | - | - |
| 60\|06 | Au NU | Small Ag seeds (22±4 nm) using 0.25 mM tannic acid with PVP in the gold solution | 60.1±6.8 | 6.1±1.8 | 5.8±0.8 | 1.05 |
| 61\|07 | | Medium Ag seeds (23.3±3.3 nm) using 1 mM tannic acid with PVP in the gold solution. | 60.8±7.4 | 7.3±2.4 | 8.6±2.6 | 0.85 |
| 79\|07 | | Large Ag seeds (27.1±4.1 nm) using 1 mM tannic acid with PVP in the gold solution | 79.2±4.3 | 7.3±1.4 | 7.3±1.2 | 1 |
| 50\|12 | | Small Ag seeds(22±4 nm) using 0.25 mM tannic acid | 50±8.8 | 12±4.4 | 6.4±1.3 | 1.9 |
| 55\|13 | | Medium Ag seeds(23.3±3.3 nm) using 1 mM tannic acid | 55±6.8 | 13±5.7 | 7.6±1.9 | 1.7 |
| 65\|15 | | Large Ag seeds(27.1±4.1 nm) using 1 mM tannic acid | 64.8±16.9 | 15.3±5.4 | 10.1±1.8 | 1.5 |

Three-dimensional visualization of the spikey particles, the nanourchins were imaged using TEM. The imaging processed sequence compiled into a smooth, stabilized video enabling clear visualization of the 3D organization of the particles is available in **Video S1**. The selected frames of the video are shown in **Figure S4**.

Based on the data acquired from **Figure S4** and **Video S1** we made a digital twin of the Au NUs (**Figure S4**). Based on **Figure S4** the core is *ca.* 80±5 nm and spikes are 20±9 nm. We used 3 model types to show the influence of the spike distribution. The simplest one is the uniform model (**Figure 2**), where all spikes are of equal size. Unfortunately, it is a very simplified model and does not represent the randomness of the spikes, therefore, we created even more realistic models. For one of them, we used Sobol's sequence for spike length distribution [38,39]. It's designed to be homogeneous in distributing fluctuations, so the overall urchin shape has long and short spikes in every direction compared to the third investigated type - random spikes length distribution that can create asymmetric patterns (**Figure 2 b**). More details on the nanourchin models are provided in the supporting information. The asymmetric patterns can compensate one another for a very large population of samples like the ones we used for steady-state extinction spectroscopy or TAS measurements; therefore, the random sequence can be close to Sobol. The three proposed models with the spikes of 0 (nanosphere), 6, 12, and 20 nm (**Figure 2 b**) were used as an input for EM simulations.

3.2 Experimental optical properties

The UV-Vis-nIR absorbance spectra of synthesized nanospheres and nanourchins are compared in **Figure 2**. The LSPR-related extinction peaks for Au nanospheres and Au nanourchins look quite similar, but the peaks of Au nanourchins with long spikes are much broader (**Figure 2a**). The obtained Au NU absorption spectra are similar to the reported short-spike Au nanourchins [18,40]. The long spike Au nanourchins have much broader absorption extending into the near-infrared region. This broad absorption peak looks different from the reported ones for Au nanostars and nanoflowers with comparably long spikes, indicating two absorption peaks [15,41].

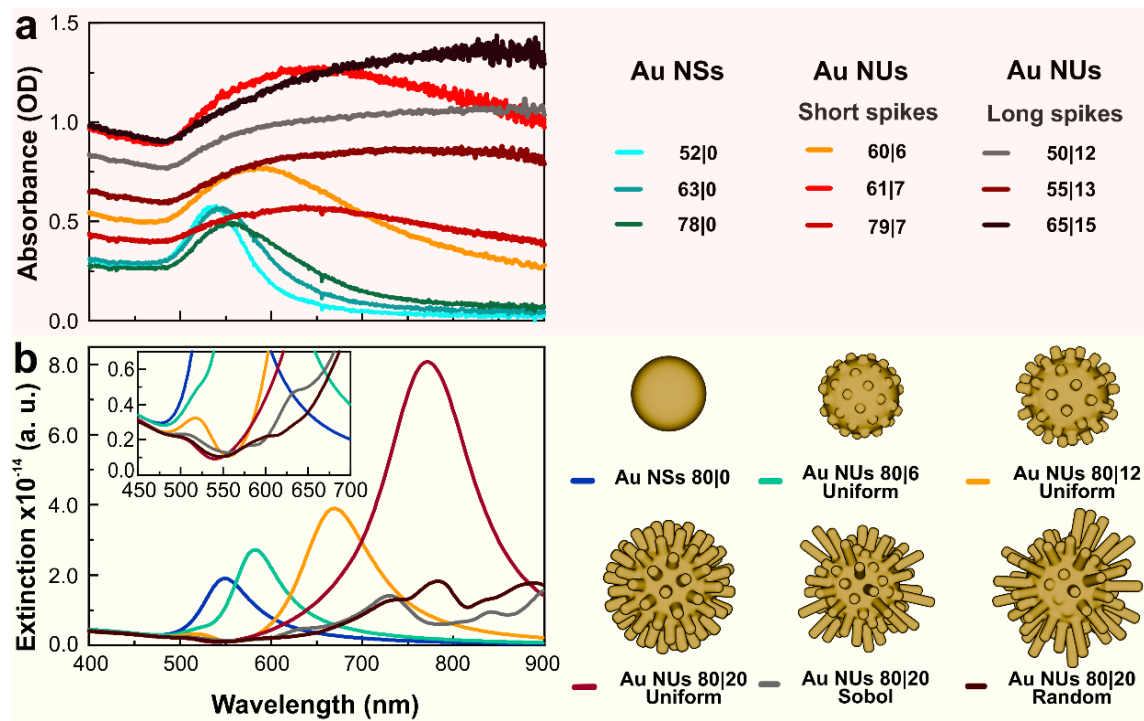

**Figure 2**. Experimental (**a**) and simulated (**b**) UV-Vis-nIR absorbance spectra of the analyzed Au spheres and nanourchins. The inset in (**b**) depicts the LSPR region related to the quadrupole of the sphere and transversal modes in the spikes. The legend of the modeling results depicts the 3D of the sphere of 80 nm diameter, uniform Au NUs of different spike lengths (6 nm, 12 nm, and 20 nm) with the same sized core, and two kinds of random Au NU models based on the TEM analysis (**Video S1**, **Figure S4**) Sobol (Sobol's sequences are a type of quasi-random low-discrepancy sequence) and random of the same average length spikes of 20 nm**.** The thickness of all spikes 10 nm.

3.3 Electromagnetic Simulation

To gain deeper insight into the optical properties of gold nanourchins and how their core and spikes influence the overall optical response, we developed a systematic EM simulation approach. Three kinds of 3D Au NU models – symmetrical, Sobol, and random, depicted in **Figure 2 b** were used, keeping the 80 nm core fixed and varying their spike lengths from 0 to 20 nm as explained in **Table S1**. Although the models were not perfectly accurate, even the representative approximate models of asymmetric nanoparticles can provide valuable information on how spike configuration changes sphere core optical and near-field distribution properties that help to explain the experimental UV-Vis-nIR absorbance spectra shown in **Figure 2 a**.

The EM simulations of a core without spikes (Au NS 80|0) resulted in a single LSPR dipole peak at 542 nm that matches the experimental UV-Vis-nIR extinction spectra and confirms the validity of the used Au dielectric function dispersions. The elongating spikes in the symmetrical spike distribution 3D model (**Figure 2 b**) resulted in the red shift of the main peak and the emergence of additional resonance at 510-520 nm (inset in **Figure 2 b**), which is mostly expressed for the 12 nm length but is visible for the 6 nm and 20 nm spike length models. The random and Sobol Au NU models also have a spectral feature at 510-520 nm wavelength,

indicating that it is related to the transversal nanorod-like LSPR of the same diameter spikes used throughout the different models. Moreover, the Mie simulations suggest that the dark quadrupolar resonance for the spheres is negligibly smaller and was not observed in the experimental extinction spectra for the Au NSs [42], but it clearly emerges in the Au NUs (uniform 80|6, 80|12).

There is a big difference between the main resonance of the uniform and different random Au NUs, despite sharing the same size 80 nm core. That indicates the significance of the random spikes, which substantially broadens the LSPR and even a single random spike Au NU depicts the experimental colloidal spectra reasonably because the same tendencies of resonance broadening are met despite differences in core size or spike length (**Figure 2**).

In the case of the uniform spikes, the well-expressed transformation from the dipolar resonance of the sphere to the far red-shifted LSPR emerges only when the aspect ratio of the spikes is over 1. The 50 shortest 6 nm length (L) and 10 nm diameter (D) spikes of 0.6 aspect ratio (L/D) expand the effective diameter of the core only by 1 nm. While the longer 12 nm length and 1.2 aspect ratio spikes or 20 nm length and 2.0 aspect ratio spikes expand the effective diameter of the core to 85 nm and 93 nm, respectively. Simplified Mie-based calculation of the gold sphere in water suggests that only a modest maximum 11.3 nm red-shift of the sphere dipolar resonance from 541.0 nm to 541.7 nm, 545.2 nm, and 552.3 nm should emerge according to the growing effective diameter (Not shown here). Although in practice, as seen in **Figure 2 b,** the modeling suggests an order of magnitude bigger shift of the peak position from the sphere (Au NS 80|0) at 542 nm to the longest spike uniform Au NUs (Au NU 80|20) at 766 nm which is 224 nm. That confirms the significance of the longitudinal nanorod-like LSPR contribution increasing along with the increasing aspect ratio and helps to explain the red-shift of the main resonance with respect to the fixed transversal resonance as was described in earlier Au nanorod studies [43–45].

To better understand the interaction of the plasmon modes, we took the 80|6, 80|12, and 80|20 nanourchins and excised the tips (80|6, 80|12, 80|20 No cores) and spherical core (80|0) and then modeled the extinction spectrum of each structure. The EM simulation results of uniform spike Au nanourchins without cores are shown **Figure S5**. The longest tips have peaks at 520 and 579 nm, while the core has a peak at 544 nm. The two plasmon bands of the 80|20 nanourchin can be explained by the nonlinear hybridization of the core (544 nm) and tip (579 nm) dipolar modes to generate a weak plasmon band at 506 nm and a very strong plasmon band at 766 nm (**Figure 3**) [46].

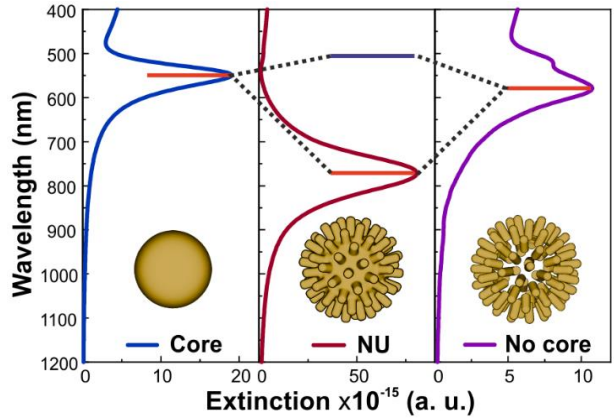

**Figure 3**. Plasmon hybridization scheme of 80|20 NU sample with a schematic energy-level diagram illustrating the plasmon hybridization mechanism in Au NUs as a result of the interaction between the spherical core and uniformly distributed spike plasmons.

The resonances and their evolution with increasing spike length are even better seen in the near-field calculations depicted in **Figure 4**. The dipolar resonance of the sphere at 542 nm (80|0 nm) shifts to the hybridized dipolar resonance at 577 nm for the 6 nm spike length Au NU (80|6 nm). Polarization maps were extracted by calculating the divergence of the electric field from the simulation data, representing the spatial distribution of induced charge density within the nanourchins under optical excitation. These maps provide insight into the nature and symmetry of the excited plasmonic modes, revealing the interplay between core and spike oscillations and identifying regions of strong local field enhancement. The electron polarization color scale helps to depict the short spike-enhanced quadrupolar resonance at 517 nm, which did not emerge clearly for a pure sphere. The uniformly distributed 12 nm spikes already hinder the core dipolar resonance at 667 nm, and the color map suggests the hybridized longitudinal resonance, while the quadrupole of the sphere at 519 nm is excited along with the transversal resonance in the spikes. The electrical field intensity clearly indicates the field concentration in the uniformly distributed spikes, which is even more expressed when they get longer. The 20 nm spikes demonstrate well-expressed transversal resonances captured at 510 nm and a hybridized longitudinal resonance peaked at 766 nm extending from the opposite sides of the NU. The transversal and longitudinal resonances at the tips are even better seen in the simulations without the core (**Figure S5**). Two absorption bands are commonly observed in Au nanostars [20,26] and nanoflowers [41] with long spikes; therefore, the optical properties modeled for Au nanourchins and their interpretation show related trends

The uniform model is easier to interpret but does not fully represent experimental observations, as theoretical resonances appear narrower than experimentally measured spectra. From **Table 1**, we can see that the spike size dispersion can be quite large; therefore, the second absorption peak position might be very broad. Therefore, Sobol and random spike distribution Au NUs were simulated with the same 20 nm average spike length on fixed 80 nm size cores. Their spectrum revealed multiple resonances (**Figure 2 b**), which are depicted in **Figure 4** and **Figure S6**. Despite random spikes, the Au NUs preserve the hybridized transversal spike resonance and quadrupolar dark mode behavior at 510 nm, as the electrical field maps do not indicate field concentration. At longer wavelengths, multiple complex resonances emerge with high field enhancement in the spikes (**Figure S6**).

The simulation findings suggest that by changing the excitation wavelength, one can either target the hybridized quadrupolar mode at 510-520 nm or using longer wavelengths at 570 nm and above, which can excite the hybridized longitudinal modes related to the bright resonances and increase the field concentration in the different length spikes. The resonances at the longer wavelengths are very sensitive to the geometry of spikes.

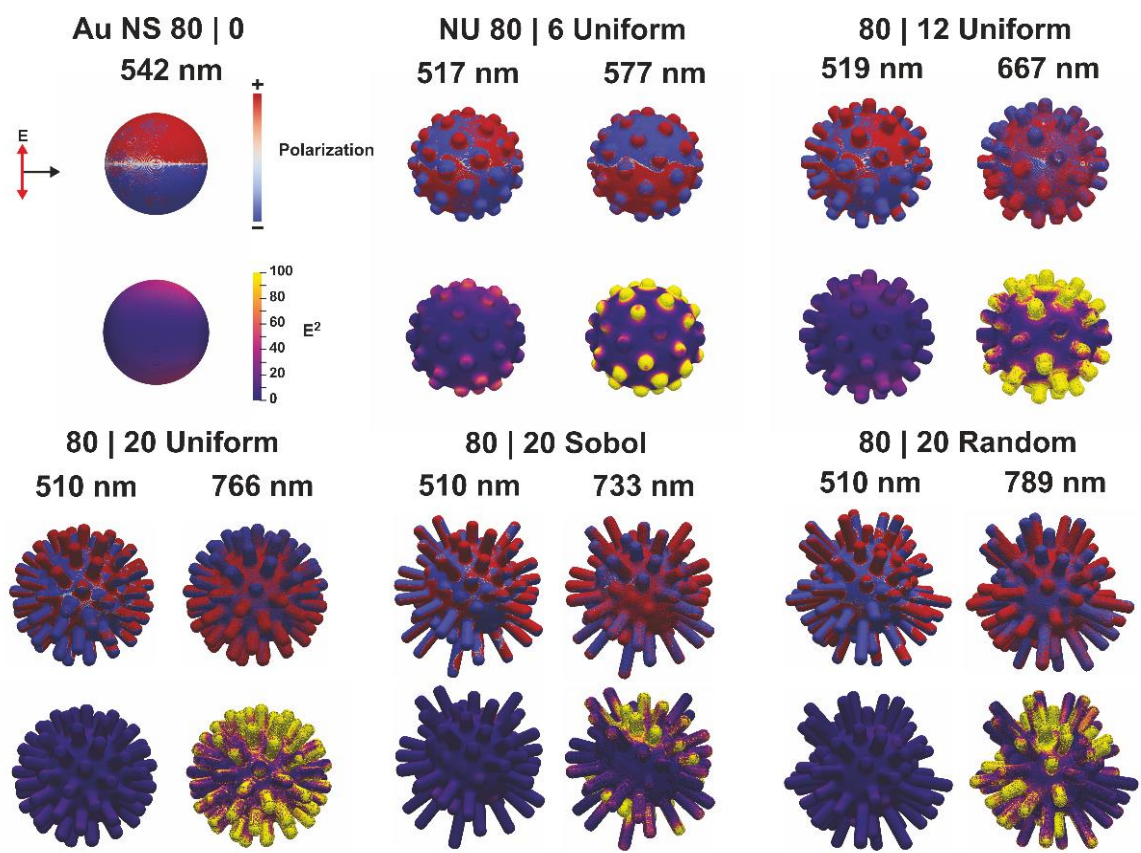

**Figure 4**. EM simulations of the polarization (above) and electric field distribution (below) for 80 nm diameter gold sphere and nanourchin structures with varying spike lengths at resonant wavelengths indicated above each set of images. The nanourchin core diameter is fixed at 80 nm, with 50 uniformly "Uniform", quasi-random "Sobol", and randomly distributed spikes "Random" having a tip diameter of 10 nm and cylindrical spikes. Spike lengths are 0 nm (sphere), 6 nm, 12 nm, and 20 nm, corresponding to Au NU samples 80|0, 80|6, 80|12, and 80|20, respectively. The color scales represent normalized polarization (top) and electric field intensity ($E^2$, bottom) for all samples.

3.4 Transient absorption spectroscopy measurement results of Au nanourchins

Aiming to elucidate the differences in the existing resonances, their spectral, and time dependencies, TAS measurements were carried out for the different core and spike size Au NUs samples summarized in **Table 1** that were excited at a broad range of wavelengths from UV to NIR (350, 500, 600, 700, 750 and 800 nm) aiming to individually excite different nature complex LSPR modes present in Au NUs. The original spectral dependencies and time traces for all investigated samples can be found in the supporting information **Figures S7-S15**. The LSPR decay times obtained were in the range of 1-2 ps, which is characteristic for gold nanostructures [11], meaning that the Ag seeds present in the core do not affect the plasmonic properties of Au NUs significantly. However, a clear dependence on the Au NU linear dimensions on pump wavelength was not observed for electron-phonon (*e-ph*) coupling times. The *e-ph* coupling times of Au NUs are quite similar to Au NSs. Whereas spectral dependencies, especially when normalized at 0 ps delay time to the amplitude of the first negative peak to -1 in the spectra as depicted in **Figure 5** had expressed differences depending on the applied pump wavelengths and length of the Au NU spikes. Short 6-7 nm length spike samples (**Figure 5 d, e, f**) indicated nearly identical TAS spectra despite the used pump wavelength, similar to the Au NS samples of a similar diameter (**Figure 5 a, b, c**) representing the Au NU cores. In contrast, the long spike 12-15 nm Au NUs had a clear difference in the amplitude of the negative TAS signal in the 600-800 nm probe band, depending on the used pump wavelength. The traces exited under $350 \leq \lambda_{pump} \leq 600$ nm had almost 0.5 a.u. the difference in amplitude (indicated by an arrow in **Figure 5 g, h, i**) compared to excited with 700 nm $\leq \lambda_{pump} \leq$ 800 nm. This finding confirms the hypothesis that longer spike lengths in Au NUs lead to a stronger hybridized nanorod-like longitudinal LSPR contribution and broader resonances, as evidenced by both electromagnetic simulations and experimental optical extinction spectra (**Figure 2**). The combined interpretation of the optical simulations and transient absorption measurements further validates the ability to synthesize Au NUs with controlled average spike lengths and to spectrally tune their field enhancements, particularly when spike lengths exceed 10 nm.

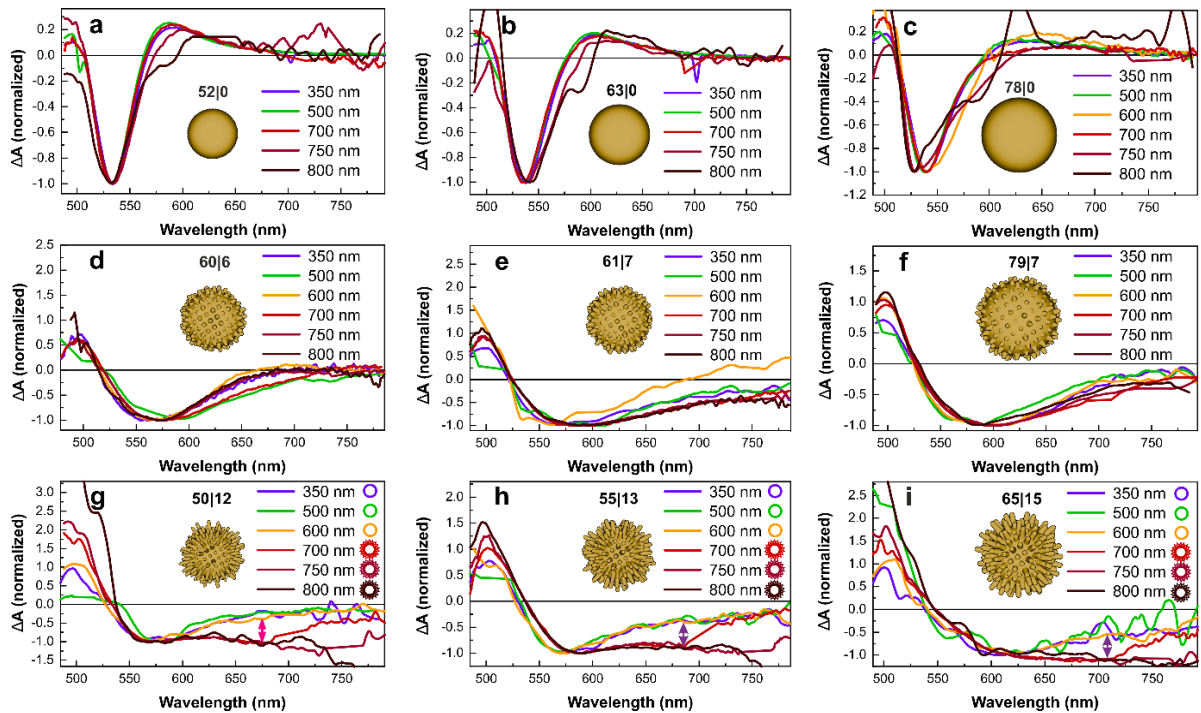

**Figure 5**. Normalized to -1 negative TAS spectra of samples 52|0 (**a**), 63|0 (**b**), 78|0 (**c**), 60|6 (**d**), 61|7 (**e**), 79|7 (**f**), 50|12 (**g**), 55|13 (**h**), and 65|15 (**i**) at a delay time of 0 ps. The excitation wavelength is depicted in the legend. The vertical double arrow in (**g-i**) indicates the change in TAS spectra at specific excitation. Every figure contains an inset with the representative Au NU model, which is proportional in size.

3.5 SERS measurements

Finally, building on the understanding of the resonances in Au NUs, we experimentally evaluated their Raman scattering enhancements by measuring the SERS signals of 78|0 Au NSs, short-spike 79|7 Au NUs, and long-spike 65|15 Au NUs at two excitation wavelengths, 532 nm and 785 nm. These wavelengths were selected to reasonably overlap with the quadrupolar resonances observed in the short-spike structures and the hybridized longitudinal resonances observed in the long-spike nanourchins (**Figure 2**, **Figure 3**). SERS measurements were performed using nanoparticle monolayers deposited on silicon substrates. Layer quality varied as seen from SEM micrographs (**Figure S16**). It was observed that spherical nanoparticles formed a better quality densely packed layer compared to non-continuous deposition for nanourchins' counterparts, which might have impacted the Raman signal enhancement. The analysis with model $10^{-4}$ M concentration 2-naphthalene thiol (2NT) molecule (representative Raman scattering spectra of the molecule are shown in **Figure S17**) revealed that the highest signal quality was observed for sample 78|0, both excited with 532 nm and 785 nm wavelength (**Figure 6**). This result corresponds well with observations by others if a monolayer of spheres can be treated as aggregated particles [47]. The SERS signal is strongly dependent on the surface structure and material used for the signal enhancement. It was previously shown that different gold structures can be optimized for 633 nm [48] or 785 nm [47] wavelength excitation, but less or no signal enhancement is registered using 532 nm excitation [47,48]. Our results show that we could detect considerable enhancement with 532 nm excitation in the case of spherical particles, but a higher signal was detected with 785 nm excitation, which could confirm the previous observation that gold is showing higher signal enhancement using excitation wavelengths red-shifted from the LSPR peak [49]. Comparing samples, the lower enhancement for the nanourchins compared to nanospheres could be partially attributed to the poorer surface packing of nanourchins. The quality of surface packaging directly corresponds to the hot spot surface density that is responsible for higher signal counts. For the sample 65|15 it was possible to identify four peaks attributable to 2NT molecule (767 $cm^{-1}$, 846 $cm^{-1}$, 1066 $cm^{-1}$, 1380 $cm^{-1}$), whereas for the short spike sample, only two peaks (767 $cm^{-1}$, 1380 $cm^{-1}$, **Table S2**) were discernible. The excitation with 785 nm resulted in higher intensity and better quality signals from all the samples in general. Longer spike urchins outperformed shorter spike counterparts with at least 2 times higher signal intensity. We have calculated the signal enhancement factor (EF, equation (S2) [50] for long spike sample comparing the signal of $10^{-2}$ M 2NT molecule measured without nanoparticles (**Figure S17**). The equation used for calculation is provided in the supporting information. Calculated enhancement factors for two peaks at 767 $cm^{-1}$ and 1380 $cm^{-1}$ using 785 nm excitation for 78|0, 79|7, and 65|15 samples are summarized in **Table 2**. It is visible that EF for long spike nanourhins is *ca*. 6-fold higher than for short spike nanourchins, which supports the hypothesis that the hybridized LSP resonance and field concentrations in the spikes observed in the EM simulations are contributing as the enhancement factor (**Figure 6** inset). This SERS investigation is more of exploratory research rather than an in-depth study, and further experiments would be necessary, but even in the present form it already signifies the importance of the nanourchins geometries and spike lengths for the electromagnetic enhancement effects and reveals the potential of nanourchins for SERS.

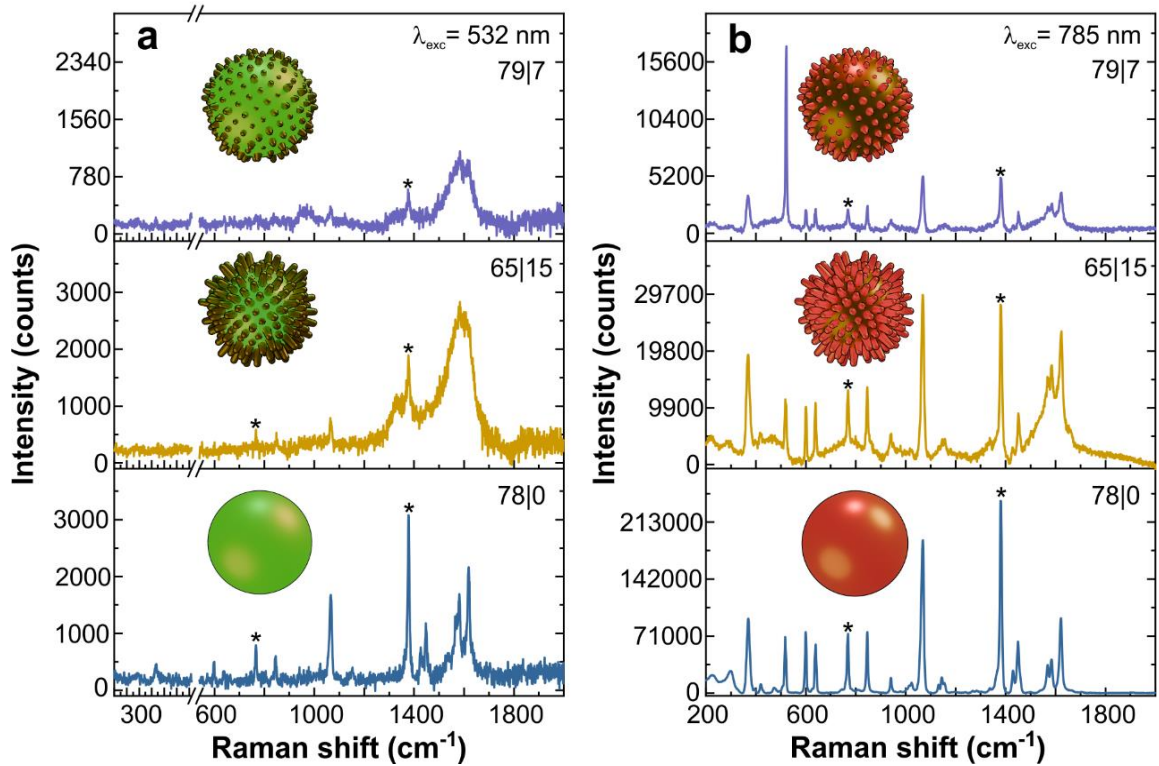

**Figure 6**. Surface-enhanced Raman-scattering spectroscopy data of Au NP samples 78|0; 79|7, and 65|15 under excitation at 532 nm (**a**) and 785 nm (**b**). The asterisk indicates molecular vibration bands used for the EF analysis. Insets depict simplified visualization of spike and core resonances characteristic for Au NUs and spheres under different wavelength excitations.

**Table 2**. SERS enhancement factor (EF) for Au NS and Au NU samples evaluated for the two characteristic bands identified in **Figure 6** and the EF ratio between short spike and long spike samples.

| Peak position, cm$^{-1}$ | EF | | | |
|---|---|---|---|---|
| | 78|0 | 65|15 | 79|7 | Ratio [EF$_{65|15}$/EF$_{79|7}$] |
| 767 | $1.5 \times 10^4$ | $0.27 \times 10^4$ | $4.1 \times 10^2$ | 6.6 |
| 1380 | $1.3 \times 10^5$ | $0.16 \times 10^5$ | $2.8 \times 10^3$ | 5.7 |

## 4. Conclusion

The study confirmed that the proposed wet synthesis method is capable of delivering controlled spike length Au NUs), which drives size-dependent hybridized LSPR resonances having strong electrical field enhancements > 600 nm.

The EM simulations suggest that spike lengths tend to override the core dipole moment seen only for ≤ 1 aspect ratio spikes, which are dominated by the hybridized longitudinal resonances typical for nanorods. The transverse resonance of the spikes hybridizes with the quadrupolar resonance of the spherical core.

High aspect ratio Au NU spikes (aspect ratio > 1) deliver strong electric field enhancements concentrated at the spike tips. Uniform Au NU models are good for interpreting the significance of the elongating spikes, but the proposed random Au NUs with the same average spike length delivered much closer spectra to the experimental ones, in spite of the use of a single Au NU in the modeling and size distribution in the colloidal sample.

The time-resolved transient absorption spectroscopy analysis suggested that the applied pump wavelength had a strong effect on the overlap with the spike-related resonances met at 700-800 nm while the LSPR decays did not indicate linear dimension-related dependence, suggesting uniform material qualities.

The wavelength-dependent Raman enhancement of the in-resonance and out-of-resonance Au NUs with different length spikes confirmed the findings in the simulations and TAS, highlighting the significance of spike lengths greater than 10 nm for creating hybridized longitudinal resonances responsible for the strong confinement of the electrical field in the spikes.

This study demonstrates that the plasmonic and optical properties of Au nanourchins strongly depend on the length and aspect ratio of their spikes, which could be beneficial for practical applications such as photocatalysis, sensing, and surface-enhanced Raman spectroscopy (SERS).

**Acknowledgements**

The research was implemented within the NANOTRAACES project carried out under the M-ERA.NET 2 scheme. This project has received funding from the Research Council of Lithuania (LMTLT), agreement No MERANET-22–2.

**Conflict of Interest**

The authors declare no conflict of interest.

**Data Availability Statement**

The data that support the findings of this study are available from the corresponding author upon reasonable request.

**Table of Contents (TOC)**

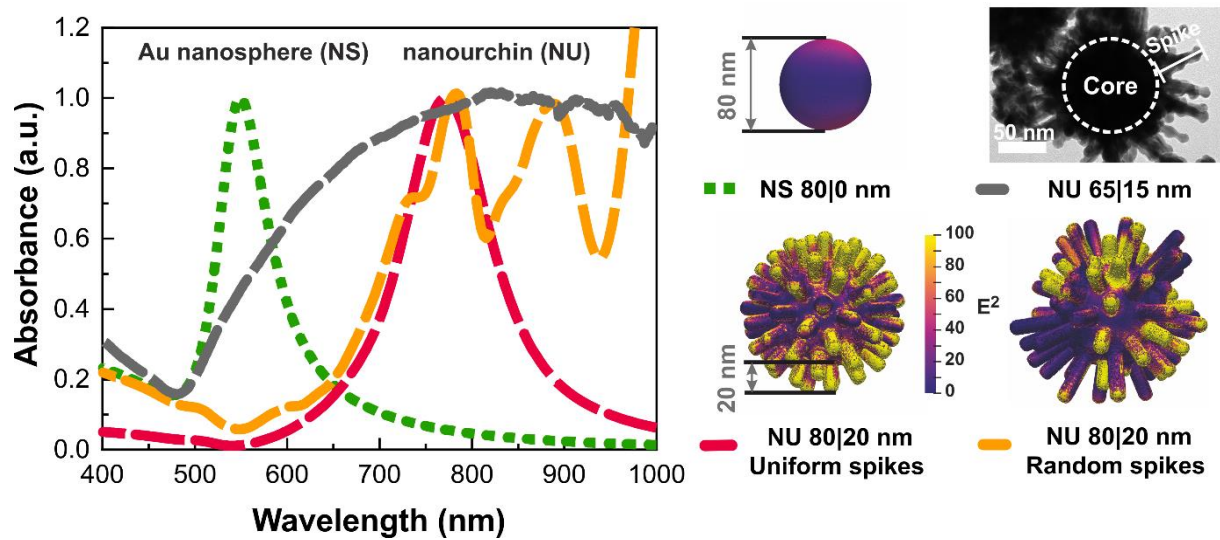

# Supporting Information

# The Effect of Spike Geometry on the Linear and Nonlinear Plasmonic Properties of Gold Nanourchins

Domantas Peckus[1,2*], Fatima Albatoul Kasabji[3,4], Maziar Moussavi[1], Loic Vidal[3,4], Hana Boubaker[3,4], Asta Tamulevičienė[1,2], Arnaud Spangenberg[3,4], Tomas Tamulevičius[1,2], Joel Henzie[5*], Karine Mougin[3,4*], Sigitas Tamulevičius[1,2*]

[1]Institute of Materials Science of Kaunas University of Technology, K. Baršausko st. 59, LT-51423, Kaunas, Lithuania

[2]Department of Physics, Kaunas University of Technology, Studentų st. 50, LT-51368, Kaunas, Lithuania

[3]Institut de Science des Matériaux de Mulhouse IS2M UMR 7361, 15 rue Jean Starcky, F 68100 Mulhouse, France

[4]Université de Strasbourg, 4 Rue Blaise Pascal, CS 90032, F-67081 Strasbourg, France

[5]International Center for Materials Nanoarchitectonics (WPI-MANA), National Institute for Materials Science (NIMS), Tsukuba, Ibaraki 305-0044, Japan

*Corresponding authors

D. Peckus: domantas.peckus@ktu.lt, orcid.org/0000-0002-4224-2521;

J. Henzie: henzie.joeladam@nims.go.jp, orcid.org/0000-0002-9190-2645;

K. Mougin: karine.mougin@uha.fr, orcid.org/0000-0002-0481-1594;

S. Tamulevičius: sigitas.tamulevicius@ktu.lt, orcid.org/0000-0002-9965-2724.

## 1.1 Chemical synthesis

### 1.1.1 Ag seeds synthesis

In the following synthesis, tannic acid serves as a reducing agent, sodium citrate plays a double role as a reducing agent and a stabilizer. The typical procedure for the production of Ag NPs has been proposed by N.G. Bastús et al. [1]. 5 mM of Sodium citrate and 0.25 mM of tannic acid were added to a 100 mL volume of aqueous solution and heated up for 15 min in a silicone oil bath in a two-neck round-bottomed flask under vigorous stirring. A spherical condenser tube was used to prevent the evaporation of the solvent. After the boiling process had started, 1 mL of $AgNO_3$ with a concentration of 25 mM was injected into the solution. The color of the solution immediately changed to yellow. After 5 min of stirring at a boiling temperature, the solution was cooled down in an ice bath to room temperature (25°C). The Ag seeds were purified by centrifugation at 14000 g for 15 min in order to remove the excess of tannic acid and then redispersed in 20 mL of ultrapure water.

### 1.1.2 Au nanourchins synthesis

The design of the reactor consisted of the "T" shaped Mixer (PEEK Tee .050 thru hole Hi Pressure F-300). The tubes (Tefzel™ (ETFE) Tubing 1/16" OD x 0.030" ID) were connected to the mixer with the adapter (Luer Adapter Female Luer to M6 Female PEEK). These three products were purchased from IDEX-HS. The lengths of the inlet and the outlet tubes were 110 cm and 164 cm, respectively. The internal diameter of the syringes was 20.2 mm, and the maximum volume of the syringe was 24 mL.

The volumes that were used are 4.5 times the previously mentioned ratio, more accurately, 4.5 mL of silver seeds synthesized with tannic acid, whose size is about 37 nm are mixed with 4.5 mL of L-DOPA and 13.05 mL of ultrapure water.

According to the formula proposed by Alessandro Silvestri et al. [2], two solutions were prepared: solution A, which is a mix of as-prepared silver seeds solution, L-DOPA (12 mM), and ultrapure water with a 1:1:2.9 ratio, while solution B is $HAuCl_4$ solution (12 mM).

Solution A and B were respectively loaded in two syringes, which were fixed in two syringe pumps. The two syringes linked the two inlet tubes of the 'T' mixer by Luer-locks, and the outlet tube was fixed on a beaker. Due to the size of the milli scale and the designed structure "T" shape of this mixer, the flow in the tube was laminar (**Figure S1**). 5 mL of Au NU solution was obtained for a flowrate ratio of 7:1 mL/min Then 12.5 mg PVP 40 was added into each Au NU solution and samples were centrifuged three times (at 10 000 g for 30 min). Ultrapure water was used to wash Au NU. Different flow rate ratios were used in order to obtain different lengths of spikes. Afterward, the Au NU samples were characterized by the UV-visible spectrometer and TEM. In order to prepare the Au NU with shorter spikes, PVP 40 was added to the $HAuCl_4$ solution before the reaction.
The NP solutions were centrifuged using an Avanti J-E JSE10L38 (Beckman Coulter Inc.) with a Rotor JA-20 (max speed 20 000 rpm or 48 400 g) that contains 8 places for a maximum tube of 50 mL.

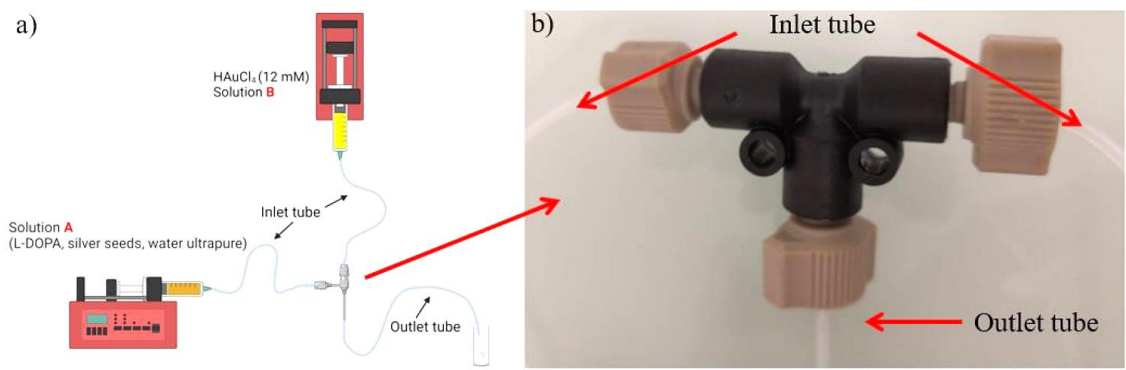

**Figure S1**. The installation illustration of the "T" shaped mixer (a); The picture of the "T" shaped mixer (b). The left and right tubes are for the inlet, and the central tube is for outlet.

In our synthesis, the role of L-DOPA is worth exploring. During the oxidation of L-DOPA, 2 hydroxyl groups in L-DOPA are deprotonated to dopa-quinone. Quinone then has an intermolecular rearrangement. After that, further oxidation occurs, and like the first oxidation, 2 electrons are gone. Afterward, another intermolecular rearrangement happens, and finally, the polymer forms via intermolecular cross-linking. Three L-DOPA molecules can react with four $AuCl_4^-$ to produce four gold atoms. The whole mechanism is shown in **Figure S2**. Moreover, this polymer produced by the oxidation of L-DOPA can form a polydopa film, which can provide protection to the formed nanoparticles from aggregation [3,4].

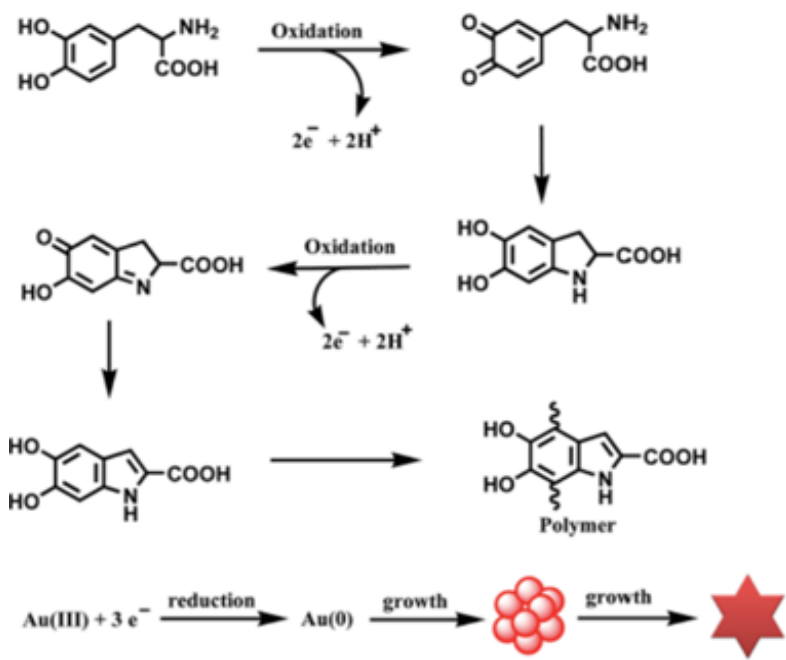

**Figure S2**. Scheme of synthesis of Au nanourchins.

The electrons released during this process can be involved in the reduction of metal cations. L-DOPA or the polymer (as shape directing agent) adsorbs on certain facets (110), (310), (720) of the initial agglomerates thus blocking further growth along these sides. Then the formation of anisotropic nanostructures occurs via expansion of the exposed facets (mainly (111)-type facets) [4,5].

Based on the results, a preliminary mechanism for crystal growth was proposed. Initially, gold nuclei were formed through DOPA reduction and rapidly grew into primary gold crystals. These primary crystals were unstable due to their high surface energy. In the next step, the primary crystals attach to the silver seeds to reduce their surface energy, resulting in the formation of branched products [6].

As the reaction progressed, additional primary gold crystals were generated and continuously adhered to the surface of the branched structures. Finally, the gold atoms in the attached crystals underwent rearrangement through crystal ripening, providing a gold source for the further development of the branched structures into well-defined gold nanourchins with more pronounced spikes [7].

## 2.1 TEM measurements

The TEM micrographs of Ag seeds used for the synthesis of Au nanourchins are depicted in **Figure S3**.

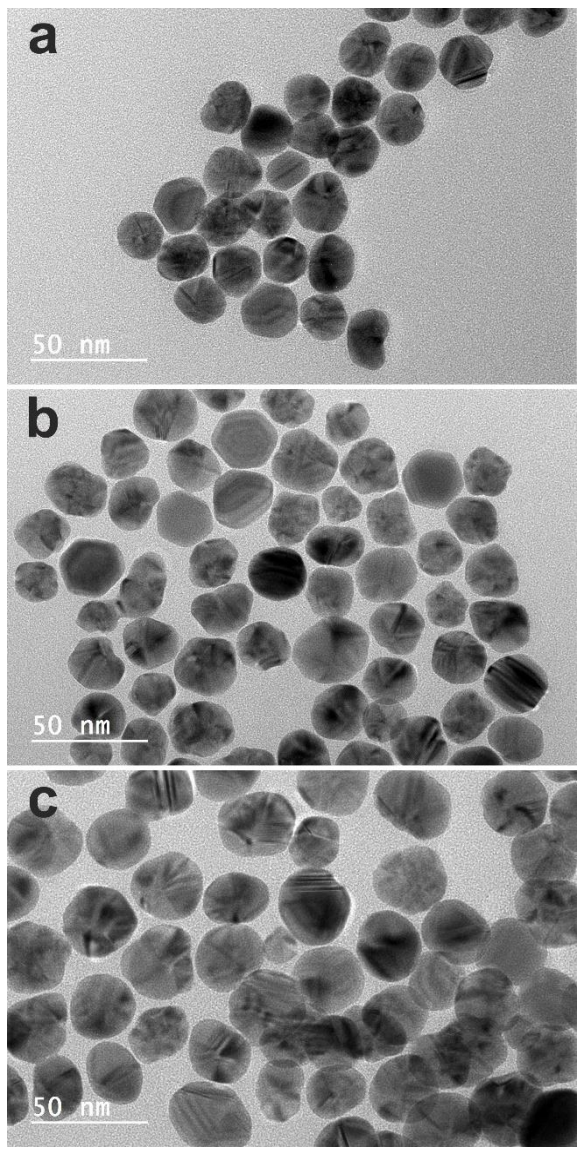

**Figure S3**. TEM images of Ag seeds used for synthesis of different Au nanourchin samples: Au NU 50|12 nm and Au NU 60|6 nm (**a**); Au NU 61|7 nm and 55|13 nm (**b**); Au NU 79|7 nm and 65|15 nm (**c**).

The sequence of differently tilted Au nanourchins used for the TEM tomographic reconstruction of the nanourchins is depicted in **Figure S4**.

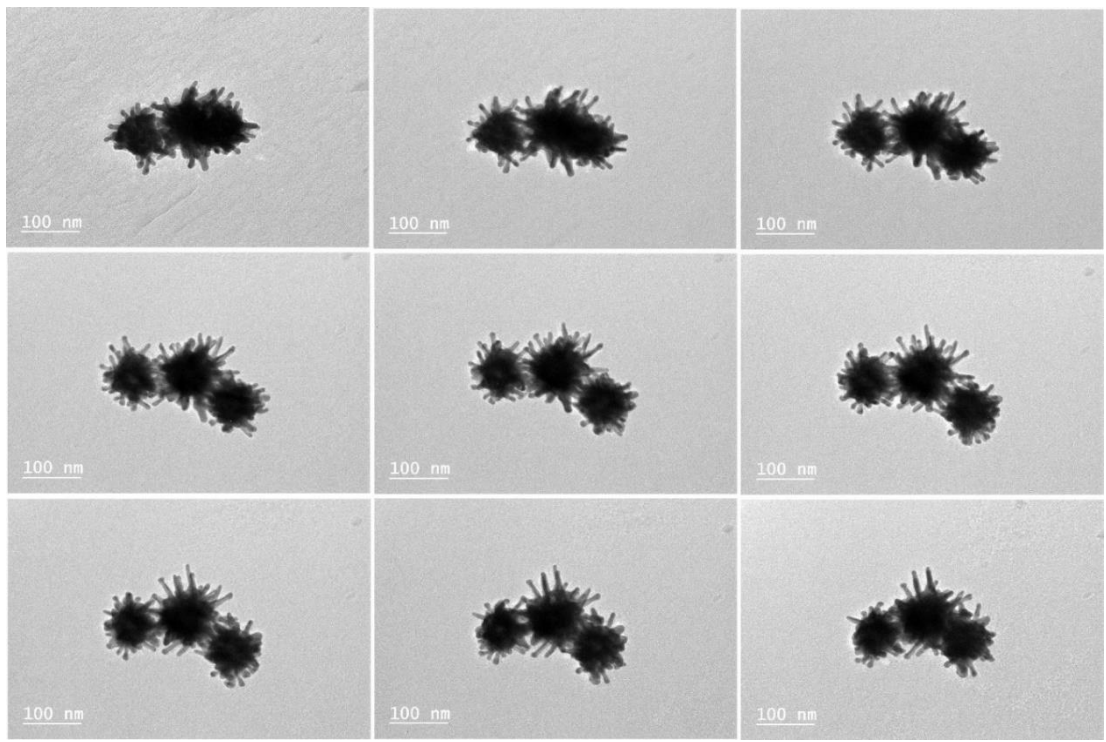

**Figure S4**. TEM images of sample Au NU 65|15 nm that were used for 3D TEM tomography and related video generation under different tilts from -50° to +60° in 1° increments There were 111 frames in total (**Video S1**).

## 2.2 Modeling of Urchins

Formulating a reproducible workflow to recreate the complex and diverse morphologies of nano-urchins is challenging. Nevertheless, by applying certain simplifying assumptions, we identified key characteristics: the Ag seed diameter, the Au core size (assumed spherical and defined by a core radius $c_r$), and spike geometry parameters such as spike length ($s_l$), spike thickness ($s_t$), and conicality vs. cylindrical shape. To parametrize conicality, we introduced a factor ($s_c$) between 0 (cylindrical spikes) and 1 (a cone with the maximum feasible angle, given the spike length, tip diameter, and base diameter).

The most crucial parameter is the number of spikes ($n_s$) and the method used to distribute them around the core. A purely random distribution can mirror certain TEM observations (**Figure 1**), but it may introduce undesirable local clustering or orientation bias, potentially skewing simulation outcomes. Random generation also complicates reproducibility. Since real experiments effectively average over many various configurations, we can either simulate a sufficiently large number of random distributions or use a more uniform approach to approximate their mean behavior. We aimed for a uniform distribution of ~50 spikes based on comparisons with TEM images (**Figure 1**, **S4**).

Achieving uniform placement of a given number of spikes over a spherical coordinate system analytically is nontrivial. While exact solutions might exist for small numbers, they become impractical for larger sets. Moreover, perfect uniformity is unnecessary for approximating real urchins. To address this, we assign each spike a polar angle ($\theta$) that sweeps from 0 to $\pi$ and an azimuthal angle ($\varphi$) defined by increments of an irrational fraction of $2\pi$. Specifically, choosing an irrational step size prevents repeated alignments and yields a more even overall distribution, even for relatively low values of $n_s$. In our case, we use the constant $\Phi = (1 + \sqrt{5})/2$ (the golden ratio) for the azimuthal increment, which helps maintain a low discrepancy pattern.

Hence, we define the orientation vector:

$$\hat{s}_i \equiv \begin{bmatrix} \theta_i \\ \phi_i \end{bmatrix} = \begin{bmatrix} \cos^{-1}\left(1 - \frac{2i}{n_s+1}\right) \\ \frac{2\pi i}{\Phi} \mod 2\pi \end{bmatrix} \quad \text{(S1)}$$

where $i$ runs from 1 to $n_s$. This approach uses a smoothly varying polar angle to cover the sphere from the "north pole" to the "south pole," while the azimuthal angle shifts by an irrational amount, preventing strong clustering along common longitudes.

A further challenge arises from the assumption that all spikes share the same length. TEM images reveal considerable variability in spike lengths, so we introduced a "spikes' fluctuation" parameter ($s_f$) ranging from 0 (no variation) to 1 (maximum variation, spanning 0 to $s_l$). Randomly assigning these lengths can again yield non-uniform clusters of very short or very long spikes, undermining reproducibility. To address this, we rely on a low-discrepancy Sobol sequence, which distributes values more evenly and prevents regional clustering. As demonstrated in **Figure 2**, a Sobol-based approach yields a more consistent spread of spikes than purely random assignment.

We consolidated these approaches into a custom MATLAB function (available on the MATLAB File Exchange, named "urchin"), which constructs 3D models of spherical urchins with user-defined particle size, spike conicality, spike-length fluctuations, and orientations. It also offers smoothing options, fast visualization, and straightforward export to 3D density masks or mesh formats.

The modeled nanourchin linear dimensions and spike parameters are summarized in **Table S1** and their EM simulations are depicted in **Figures S5, S6.**

**Table S1**. Parameters used for generating simulated models.

| Model name | Model Parameters | | | | | |
|---|---|---|---|---|---|---|
| | Core Diameter | Number of Spikes | Spikes' Length | Spikes' Fluctuation | Spikes' Conicality | Fluctuation Method |
| | $c_r$ (nm) | $n_s$ | $s_l$ (nm) | $s_f$ | $s_c$ | |
| **NS 80\|0** | 80 | 50 | 0 | 0 | 0 | - |
| **NU 80\|6 Uniform** | 80 | 50 | 6 | 0 | 0 | - |
| **NU 80\|6 Uniform No Core** | - | 50 | 6 | 0 | 0 | - |
| **NU 80\|12 Uniform** | 80 | 50 | 12 | 0 | 0 | - |
| **NU 80\|12 Uniform No Core** | - | 50 | 12 | 0 | 0 | - |
| **NU 80\|20 Uniform** | 80 | 50 | 20 | 0 | 0 | - |
| **NU 80\|20 Uniform No Core** | - | 50 | 20 | 0 | 0 | - |
| **NU 80\|20 Sobol** | 80 | 50 | 20 | 1 | 0 | Sobol |
| **NU 80\|20 Random** | 80 | 50 | 20 | 1 | 0 | Random |

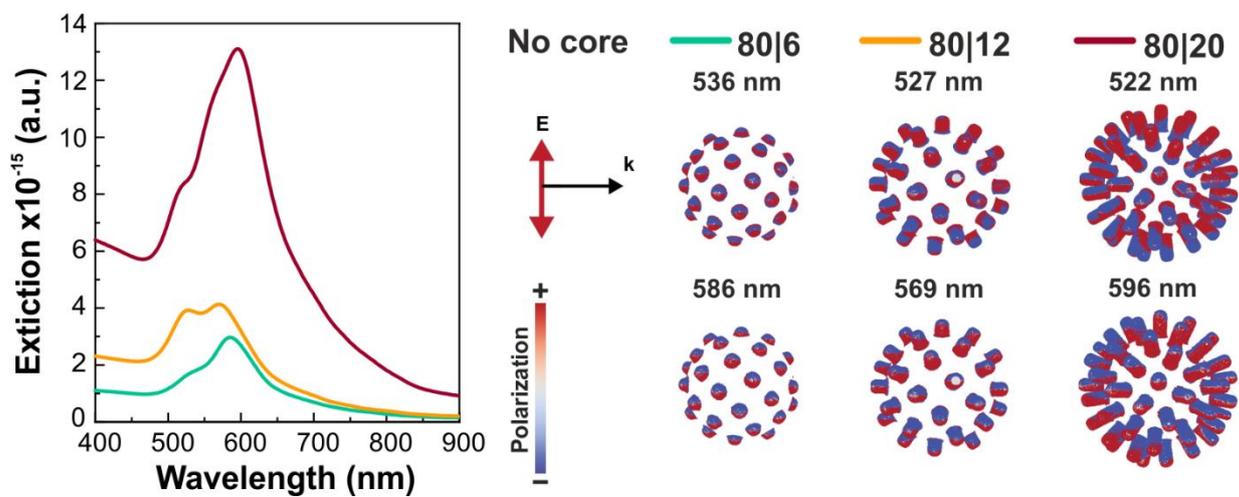

**Figure S5**. EM simulations of extinction spectra and polarization of Au nanourchins without cores of samples 80|6, 80|12, and 80|20. Polarization was calculated at peaks of extinction spectra.

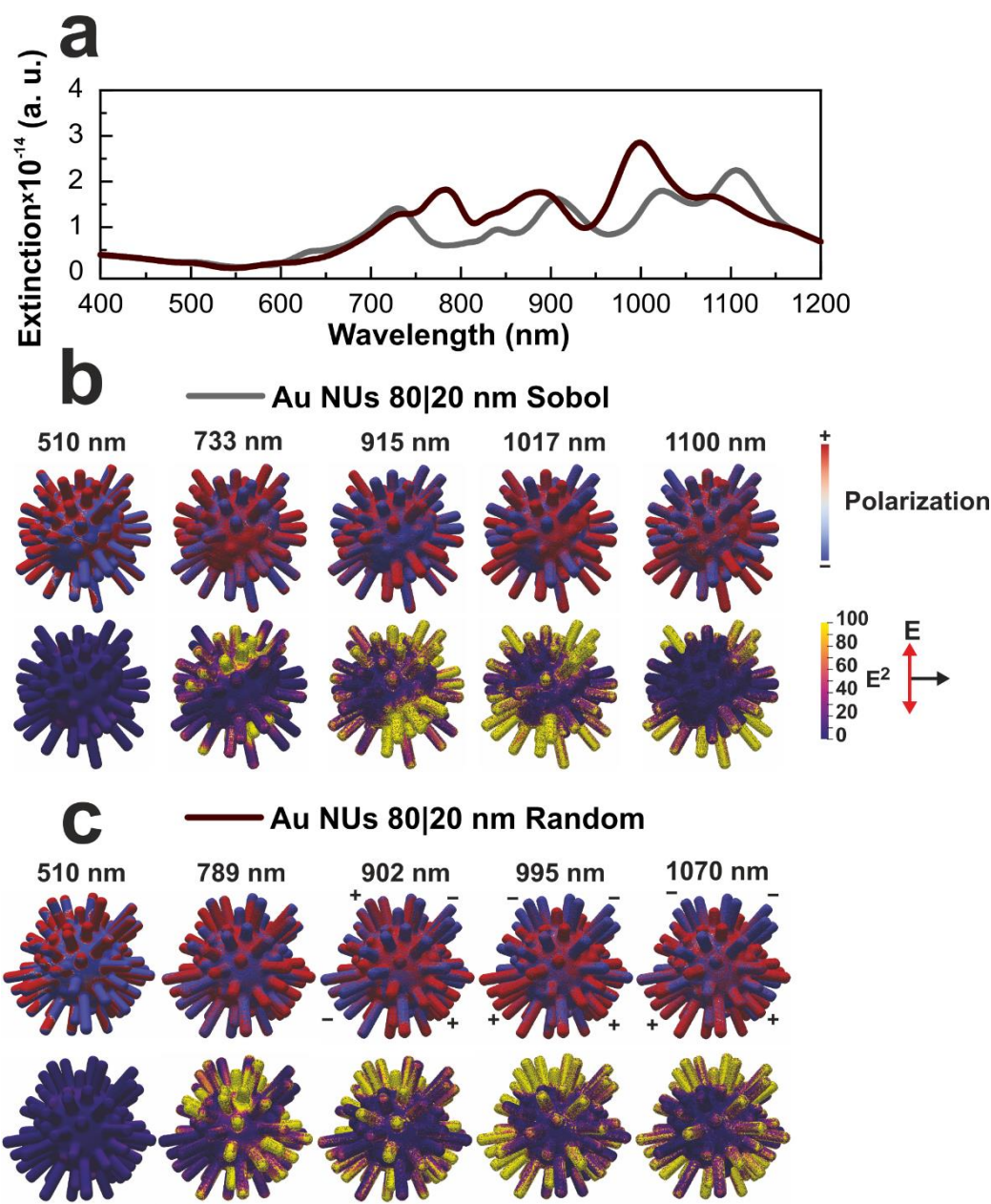

**Figure S6**. Extinction spectra of Au NU 80|20 nm Sobol (**a**) and Random models (**b**). FDTD simulations of the electric field distribution (above) and polarization (below) for 80 nm diameter nanourchin structures with varying spike lengths at resonant wavelengths indicated above each set of images. The nanourchin core diameter is fixed at 80 nm, with 50 quasi-random Sobol "Sobol", and randomly distributed spikes "Random" (**c**) having a tip diameter of 10 nm and zero conicality. Spike lengths are 20 nm. The color scales represent polarization (top) and electric field intensity ($E^2$, bottom) for all samples.

## 2.3 Transient absorption data

The TAS spectra of the Au nanosphere and nanourchin samples are depicted in **Figures S7-15**.

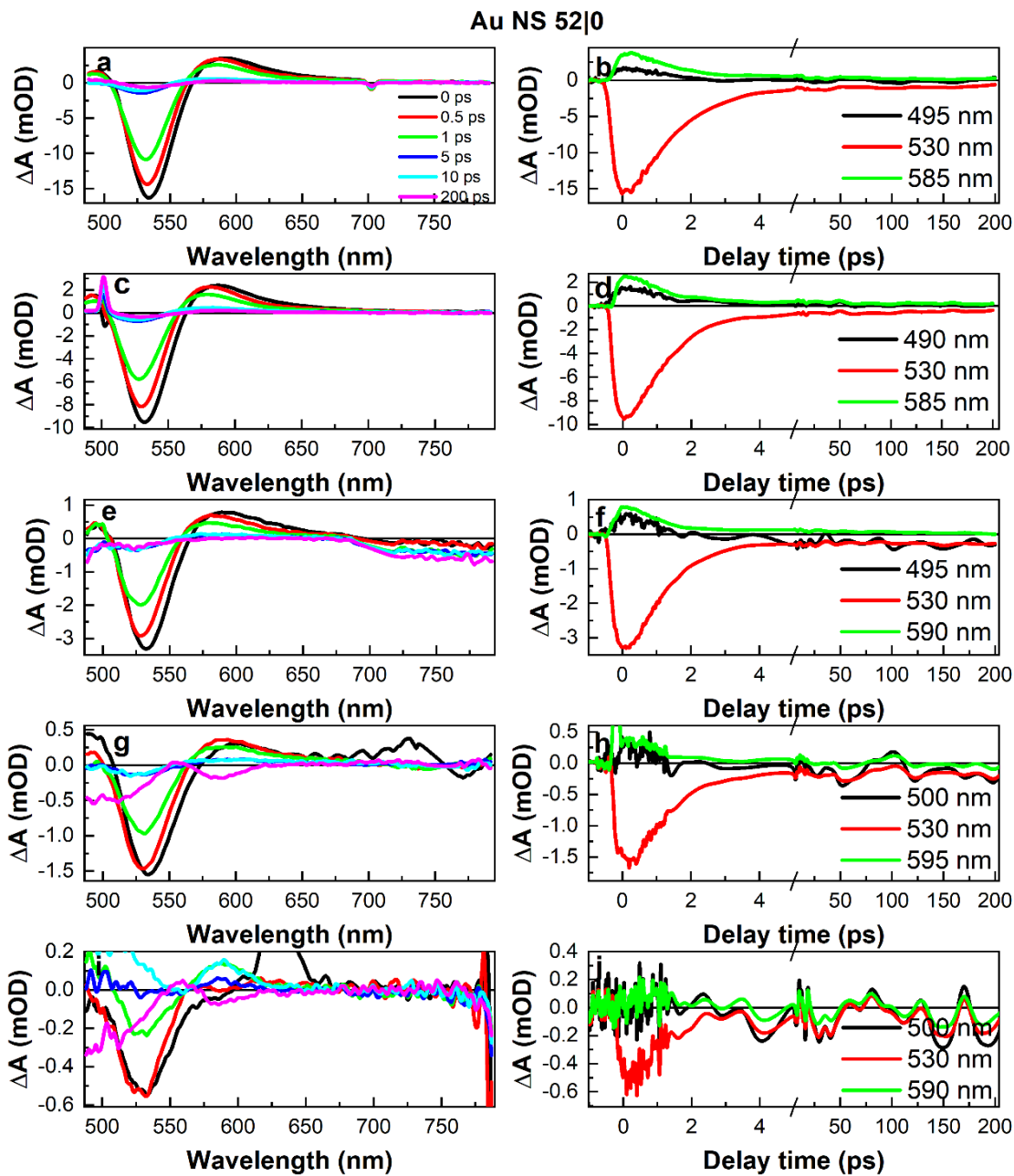

**Figure S7**. TAS spectra and traces of 52 nm diameter Au spheres (Au NS **52|0**) under excitation at 350 nm (a, b), 500 nm (c, d), 700 nm (e, f), 750 nm (g, h) and 800 nm (i, j), excitation intensities 7.8, 5.1, 57.6, 41.6, and 36 µJ/cm$^2$ respectively.

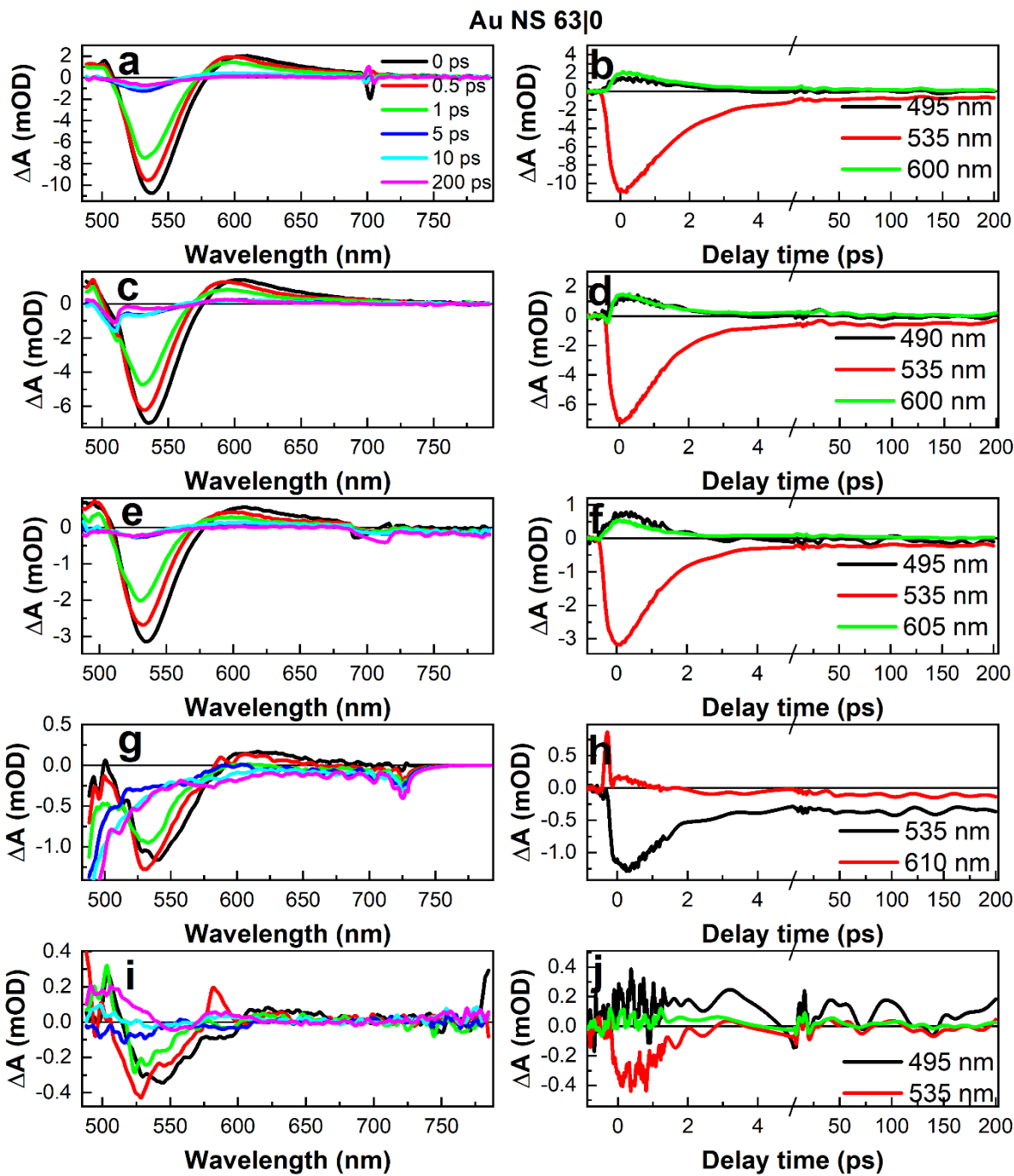

**Figure S8**. TAS spectra and traces of **62.5** nm diameter Au spheres (Au NS **63|0**) under excitation at 350 (a, b), 500 (c, d), 700 (e, f), 750 (g, h), and 800 nm (i, j), excitation intensities 7.8, 5.1, 57.6, 41.6, and 36 µJ/cm$^2$ respectively. Due to scattered excitation light at 750 nm the spectral area at 730 – 792 nm was removed for TAS spectra received under excitation at 750 nm (g).

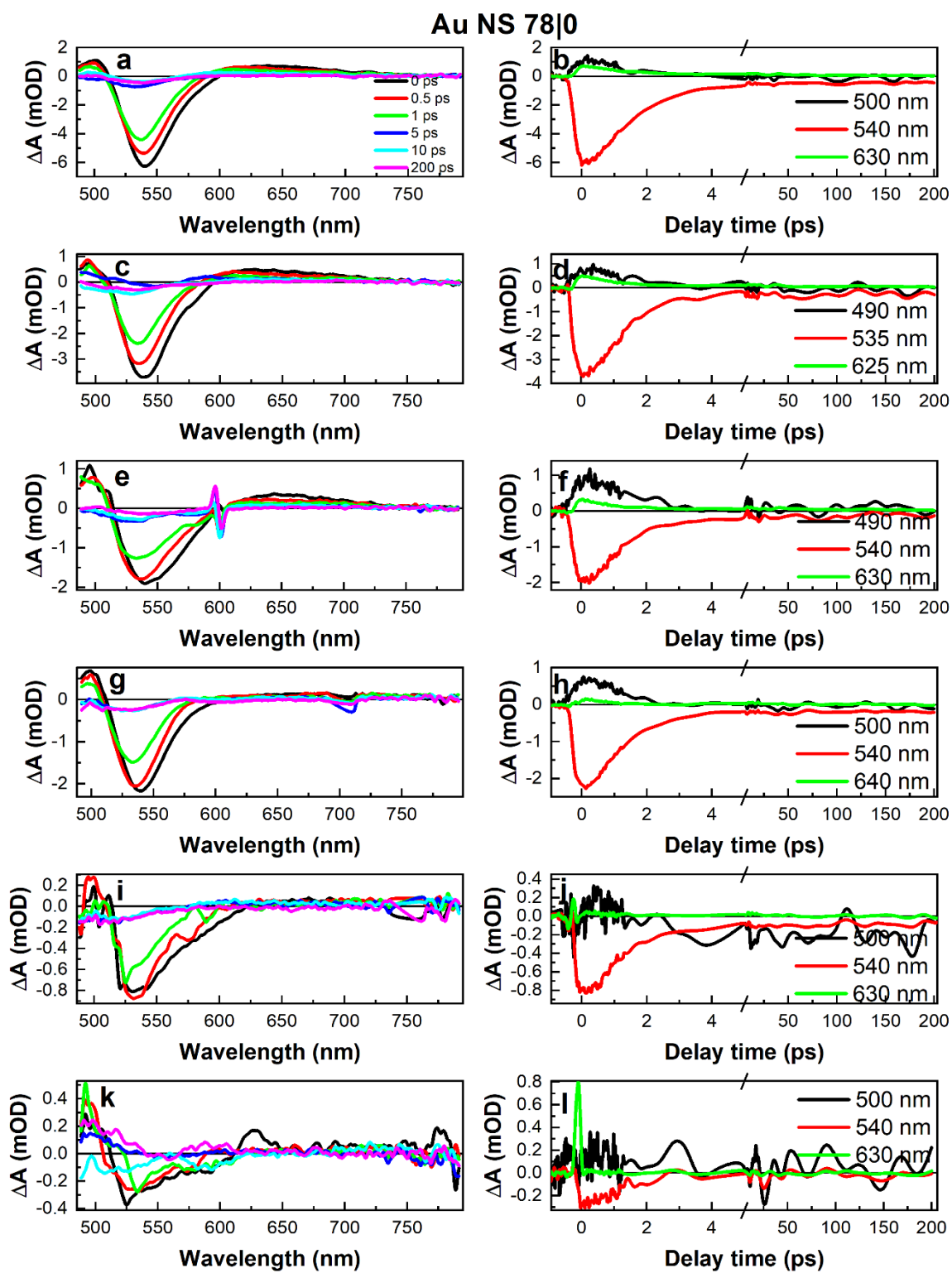

**Figure S9**. TAS spectra and traces of **77.5** nm diameter Au spheres (Au NS **78|0**)under excitation at 350 (a, b), 500 (c, d), 600 (e, f), 700 (g, h), 750 (i, j) and 800 nm (k, l), excitation intensities 7.8, 5.1, 10.4, 56.8, 41.6 and 36 µJ/cm$^2$ respectively.

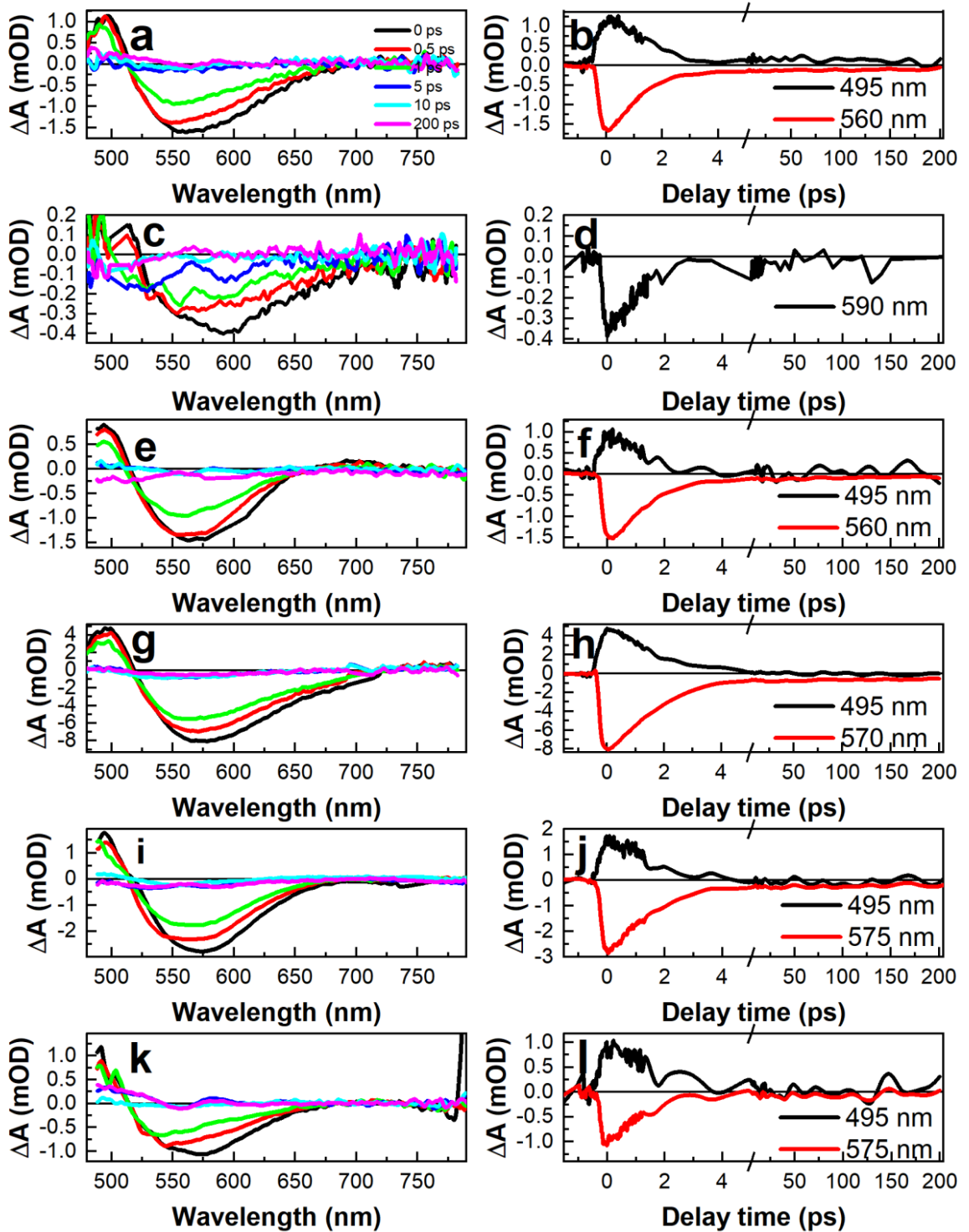

**Figure S10**. TAS spectra and traces of sample Au NU **60|6** under excitation at 350 nm (a, b), 500 nm (c, d), 600 nm (e, f), 700 nm (g, h), 750 nm (i, j), and 800 nm (k, l) excitation intensity 9.6, 2.4, 10.1, 61.6, 44, and 36 μJ/cm$^2$ respectively.

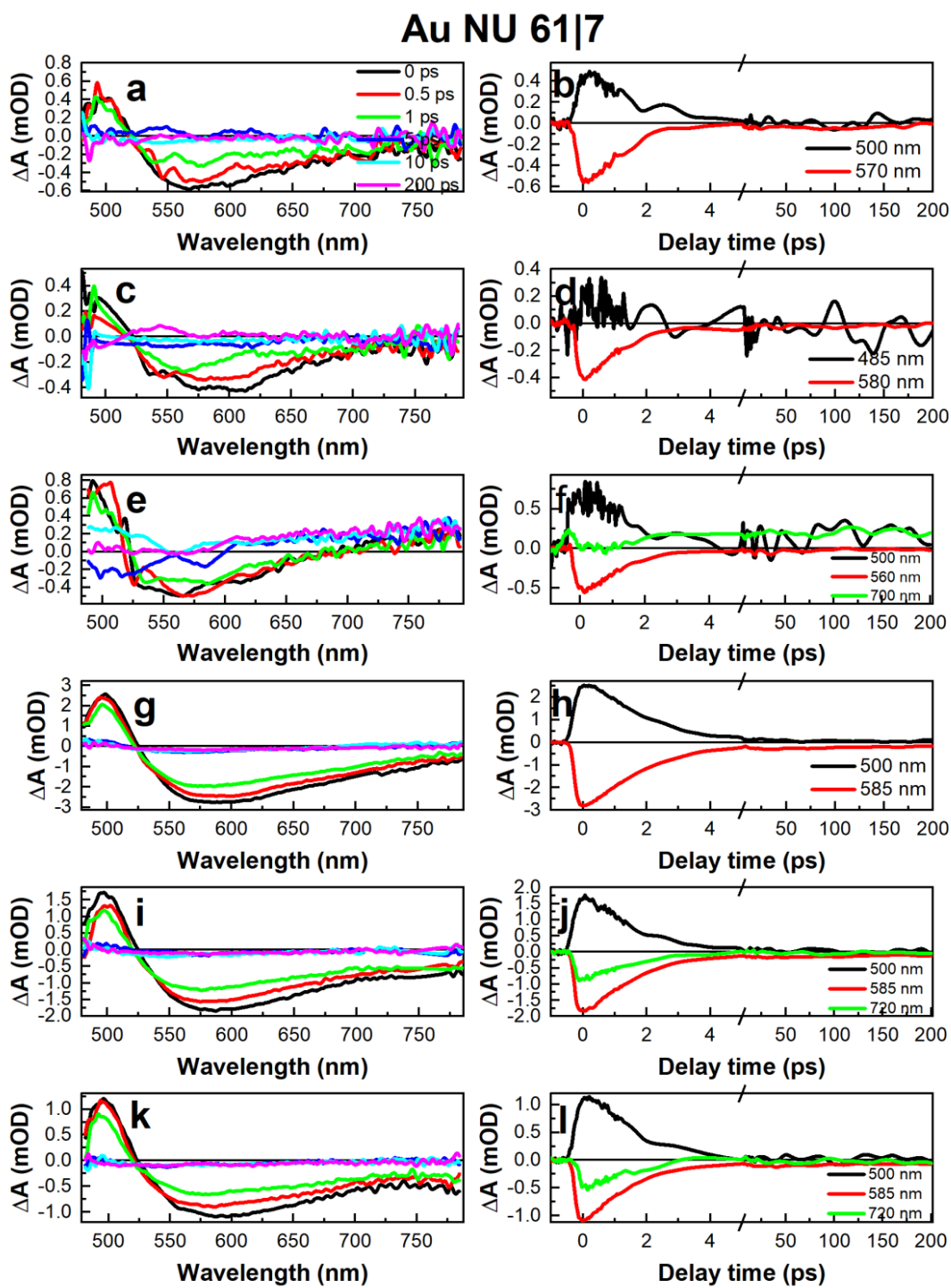

**Figure S11**. TAS spectra and traces of sample Au NU **61|7** under excitation at 350 nm (a, b) 500 nm (c, d) 600 nm (e, f), 700 nm (g, h) 750 nm (i, j), and 800 nm (k, l) excitation intensity 9.6, 4.8, 10.4, 58.4, 45.6, and 41.6 μJ/cm$^2$ respectively.

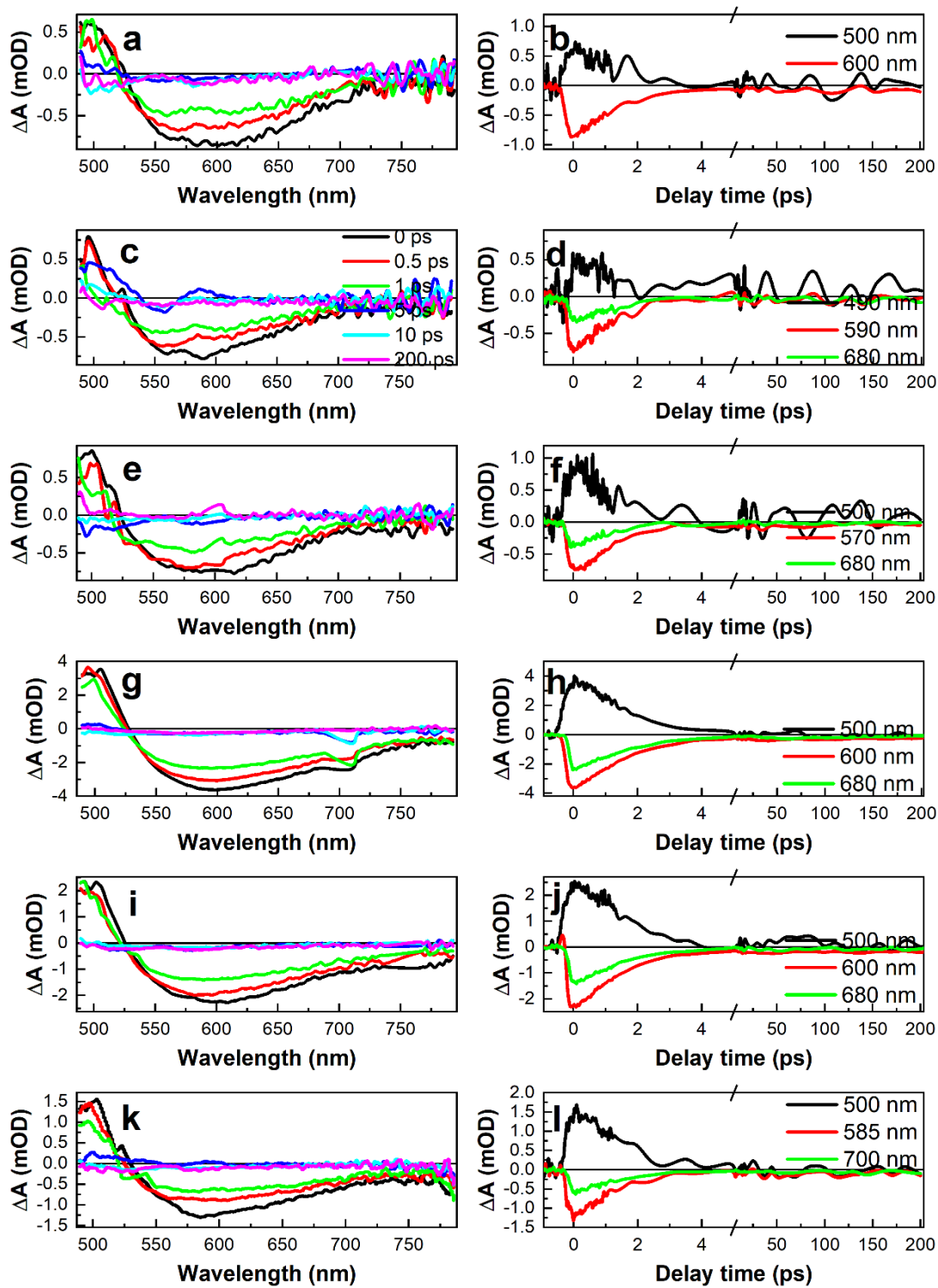

**Figure S12**. TAS spectra and traces of Au NU **79|7** under excitation at 350 nm (a, b), 500 nm (c, d), 600 nm (e, f), 700 nm (g, h), 750 nm (i, j), and 800 nm (k, l), excitation intensity 8.8, 5.6, 9.6, 56.0, 40.0, and 35.2 µJ/cm$^2$ respectively

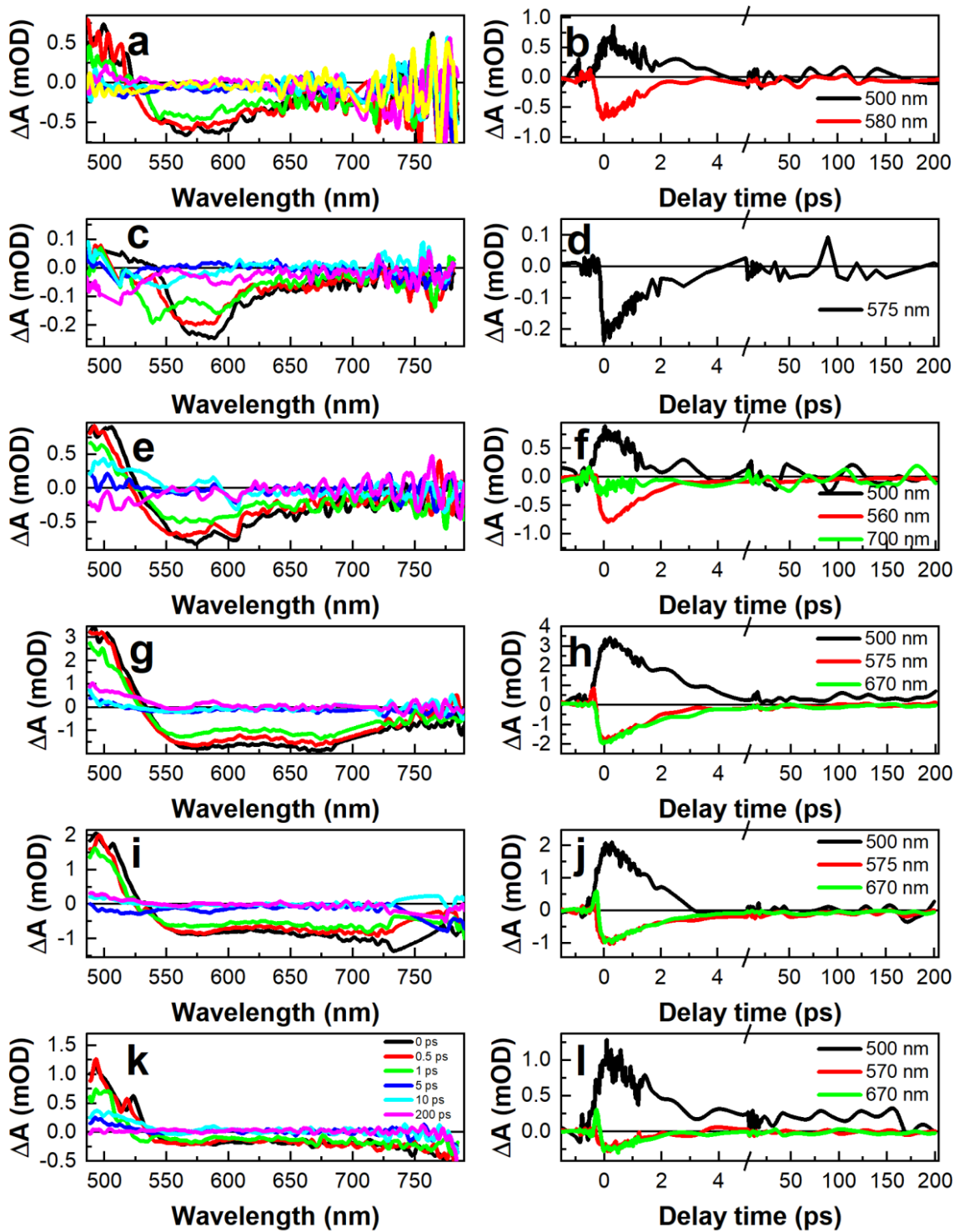

**Figure S13**. TAS spectra and traces of sample Au NU **50|12** under excitation at 350 nm (a, b), 500 nm (c, d) 600 nm (e, f), 700 nm (g, h), 750 nm (i, j), and 800 nm (k, l) excitation intensity 9.6, 2.4, 10.1, 52.8, 40.0, and 36.0 µJ/cm² respectively.

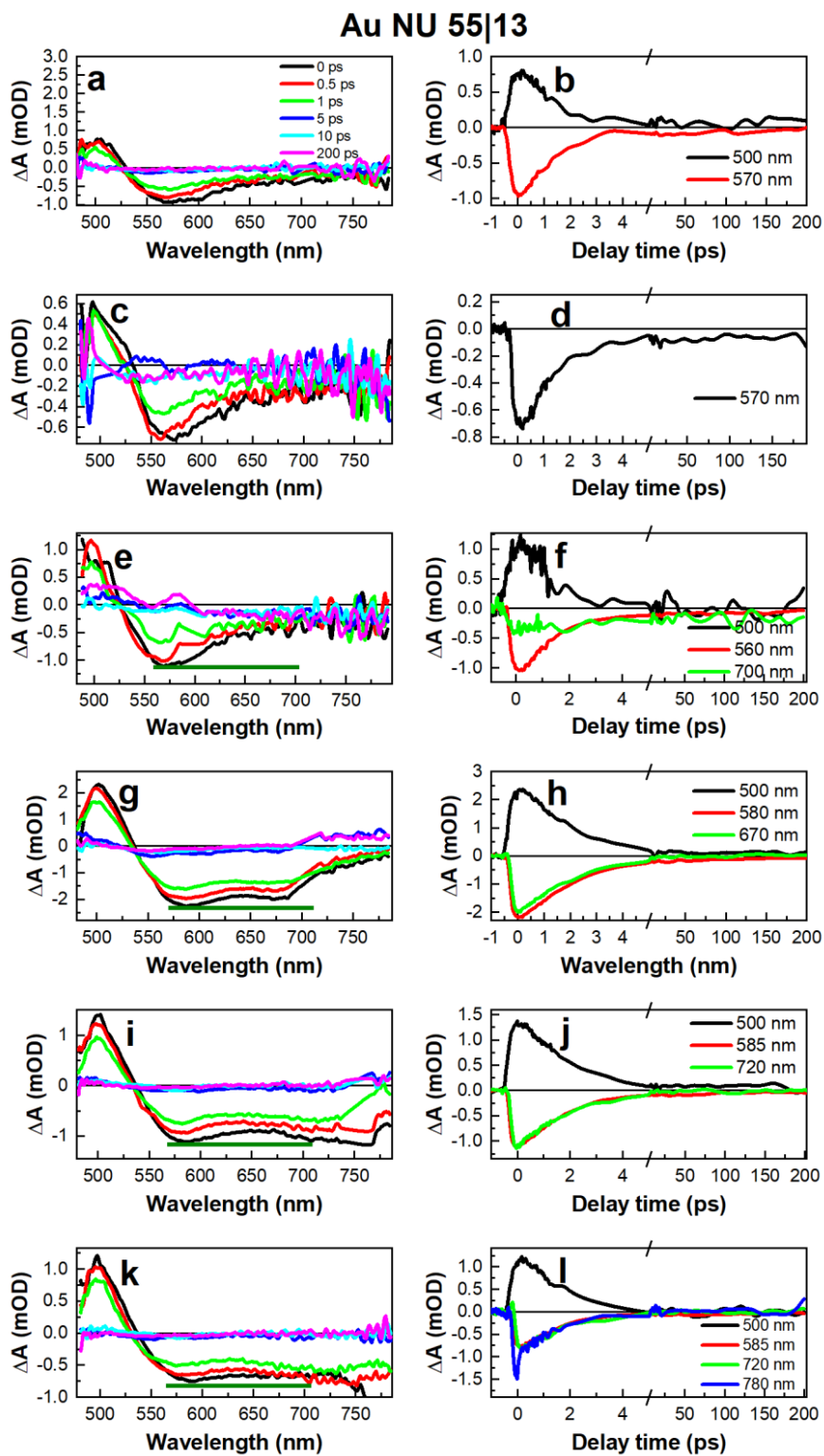

**Figure S14**. TAS absorption spectra and traces of sample Au NU **55|13** under excitation at 350 nm (a, b) 500 nm (c, d) 600 nm (e, f), 700 nm (g, h), 750 nm (i, j), and 800 nm (k, l) excitation intensity 9.6, 4.8, 10.4, 58.4, 47.2, and 41.6 μJ/cm² respectively. The **horizontal olive line** was added to the spectra figures to expose the difference in TAS spectra more clearly.

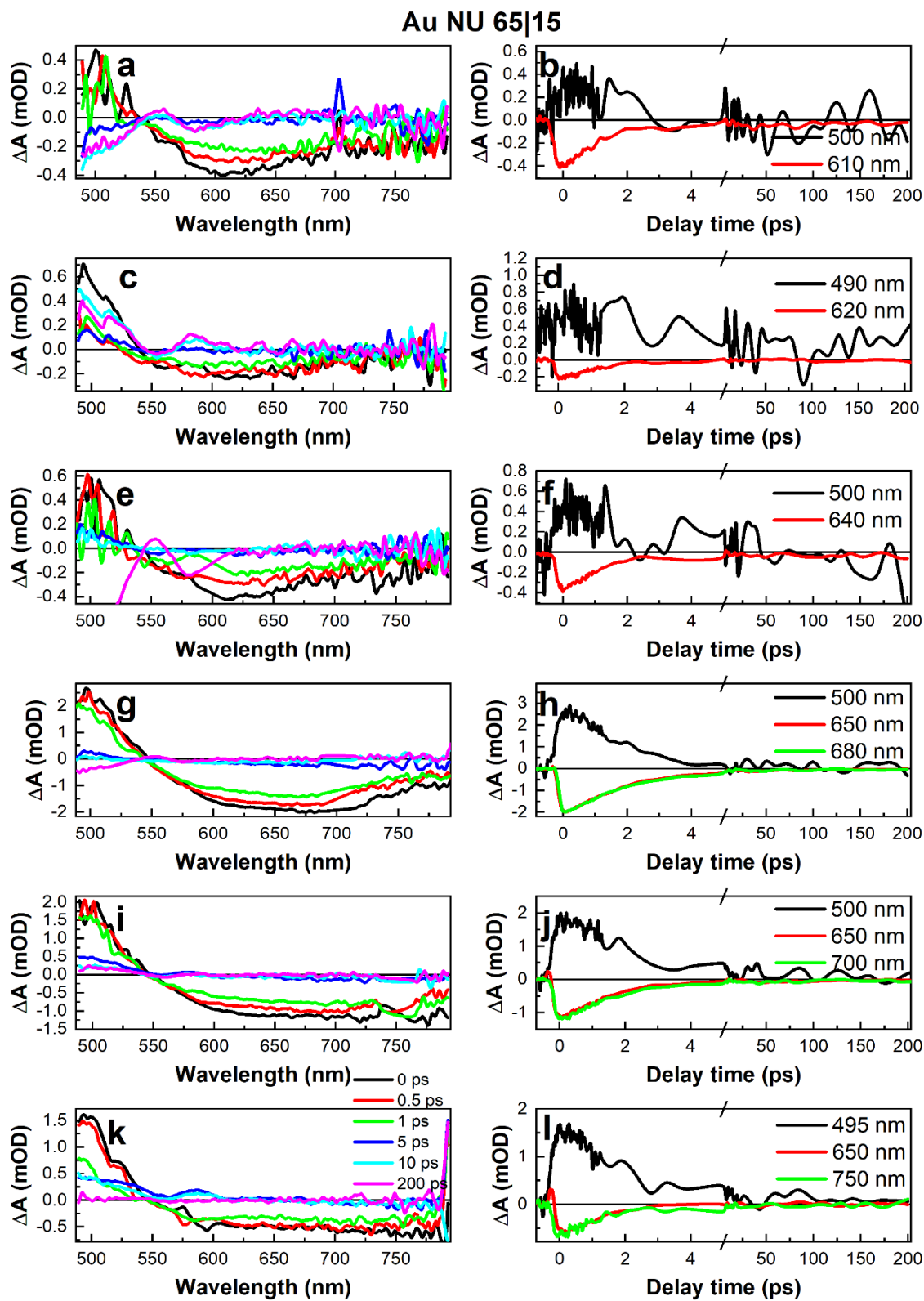

**Figure S15**. TAS spectra and traces of Au NU **65|15** under excitation at 350 nm (a, b), 500 nm (c, d), 600 nm (e, f), 700 nm (g, h), 750 nm (i, j), and 800 nm (k, l) excitation intensity 8.8, 4.8, 9.6, 56.0, 40.0, and 35.2 µJ/cm$^2$ respectively.

## 2.4 SERS measurements

The SEM micrographs of Au nanoparticle samples included in the SERS study are depicted in **Figure S16**.

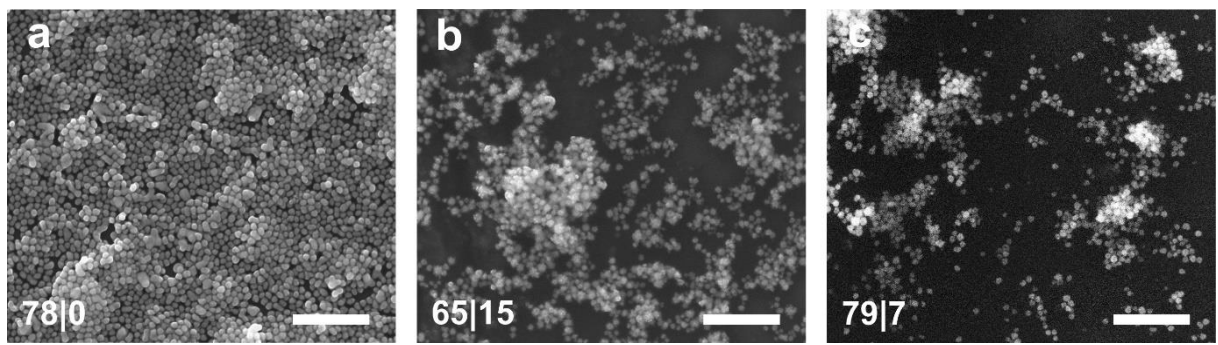

**Figure S16**. Surface morphology of monolayers used for SERS analysis: (a) nanospheres 78|0, (b) nanourchins 65|15, (c) nanourchins 79|7. The scale bar is 1mm.

### 2.4.1 Enhancement factor calculation.

Assuming that all spectra were measured under the same conditions, the analytical enhancement factor (*EF*) was calculated according to the equation [8,9]:

$$EF = \frac{I_{SERS}/C_{SERS}}{I_{Raman}/C_{Raman}} \quad (S2)$$

where, $I_{SERS}$ is the peak intensity on the sample with gold nanoparticles, $C_{SERS}$ is the concentration of the analyte at which SERS spectra were measured (in this case $10^{-4}$M), $I_{Raman}$ is the peak intensity on the sample without gold nanoparticles, $C_{Raman}$ is the concentration of the analyte ($10^{-2}$M) measured at the blank substrate.

### 2.4.2 Analyte molecule information

The Raman scattering spectra of 2-naphthalene thiol (2NT) molecules were measured with two excitation wavelengths of 532 nm and 785 nm. Material was measured in powder form and dissolved in ethanol for $10^{-2}$ M concentration. Spectra of the powder are provided in **Figure S17 a** and spectra of dissolved 2NT are provided in **Figure S17 b**. Spectra are plotted with the offset in the ordinate axis for better representation. The attribution of the peaks is given in **Table S2** according to [10].

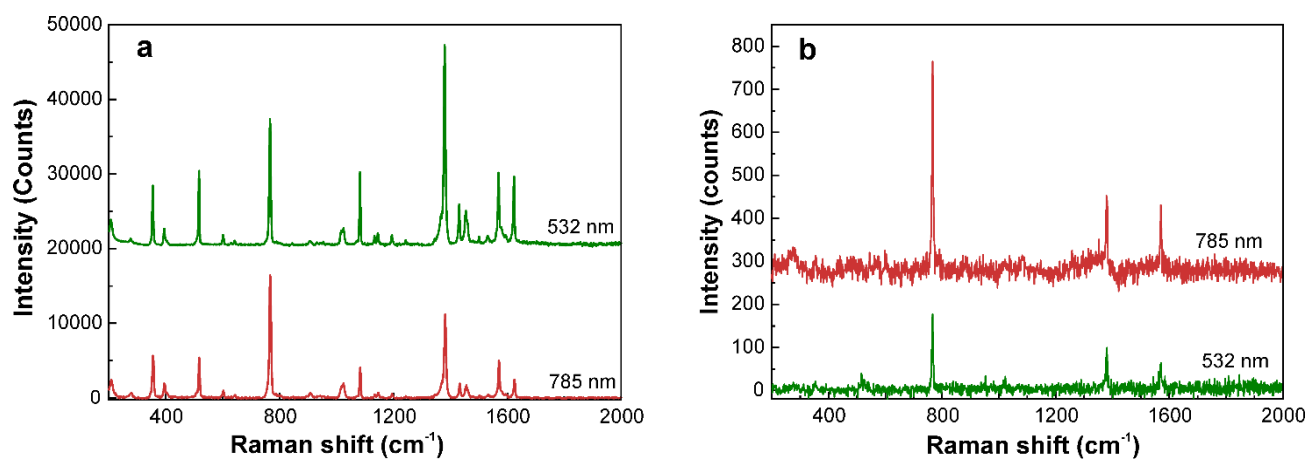

**Figure S17**. Surface-enhanced Raman-scattering spectroscopy data of powder (a) and solvent (b) of 2NT.

**Table S2**. Peaks in the Raman scattering spectrum of the 2NT molecule [10].

| Peak position (cm$^{-1}$) | Assignment |
|---|---|
| 354 | Ring deformation + C–S |
| 517 | Ring deformation |
| 767 | Ring deformation |
| 1082 | C-H bend |
| 1380 | Ring stretch |
| 1431 | Ring stretch |
| 1569 | Ring stretch |
| 1623 | Ring stretch |

# Referencies